\pdfoutput=1
\documentclass[aps, reprint, superscriptaddress, twocolumn,longbibliography,nofootinbib,floatfix]{revtex4-2}
\usepackage[utf8]{inputenc}

\usepackage[T1]{fontenc}

\usepackage{graphicx}
\usepackage{textcomp}
\usepackage{tabularx} 
\makeatletter
\renewcommand*{\@fnsymbol}[1]{\ensuremath{\ifcase#1\or *\or**\or \dagger\or \ddagger\or
   \mathsection\or \mathparagraph\or \|\or  \dagger\dagger
   \or \ddagger\ddagger \else\@ctrerr\fi}}
\makeatother
\usepackage{fancyhdr}            
\fancypagestyle{plain}{
    \lhead{}
    \fancyhead[R]{\thepage}
    \fancyhead[L]{}
    
    \fancyfoot{}
}
\usepackage[columnwise,modulo]{lineno}

\usepackage{amsmath,amssymb,amstext,graphicx,bm,booktabs}
\usepackage[hidelinks]{hyperref}
\usepackage{fancyhdr}
\usepackage{microtype}
\usepackage{bm}
\usepackage{xcolor}
\usepackage{multirow}
\usepackage{placeins}
\usepackage[font=small,labelfont=bf,
   justification=centerlast
   ,
   format=plain]{caption}
\usepackage{subcaption}
\usepackage{ragged2e}
\newcommand{\justifiedcaption}[1]{%
  \refstepcounter{figure}%
  \centering
  \vspace{5pt}
  \begin{minipage}{\linewidth}
    \textbf{FIG.~\thefigure:} \justifying#1\par
  \end{minipage}
}
\newcommand{\justifiedtablecaption}[1]{%
  \refstepcounter{table}%
  \centering
  \vspace{5pt}
  \begin{minipage}{\linewidth}
    \textbf{TABLE~\thetable:} \justifying#1\par
  \end{minipage}
}
\usepackage[normalem]{ulem} 
\usepackage{xcolor} 

\usepackage{soul}

\begin{document}
\title{Fully autonomous tuning of a spin qubit
}

\author{Jonas Schuff}
    \thanks{jonas.schuff@materials.ox.ac.uk}
    \affiliation{Department of Materials, University of Oxford, Oxford OX1 3PH, United Kingdom}
\author{Miguel J. Carballido}
    \affiliation{Department of Physics, University of Basel, 4056 Basel, Switzerland}    
\author{Madeleine Kotzagiannidis}
    \affiliation{Mind Foundry Ltd, Summertown, Oxford OX2 7DD, United Kingdom}
\author{Juan Carlos Calvo}
    \affiliation{Mind Foundry Ltd, Summertown, Oxford OX2 7DD, United Kingdom}
\author{Marco Caselli}
    \affiliation{Mind Foundry Ltd, Summertown, Oxford OX2 7DD, United Kingdom}
\author{Jacob Rawling}
    \affiliation{Mind Foundry Ltd, Summertown, Oxford OX2 7DD, United Kingdom}
\author{David L. Craig}
    \affiliation{Department of Materials, University of Oxford, Oxford OX1 3PH, United Kingdom}
\author{Barnaby van Straaten}
    \affiliation{Department of Materials, University of Oxford, Oxford OX1 3PH, United Kingdom}

\author{Brandon Severin}
    \affiliation{Department of Materials, University of Oxford, Oxford OX1 3PH, United Kingdom}
\author{Federico Fedele}
    \affiliation{Department of Engineering Science, University of Oxford, Oxford OX1 3PJ, United Kingdom}
\author{Simon Svab}
    \affiliation{Department of Physics, University of Basel, 4056 Basel, Switzerland} 
\author{Pierre Chevalier Kwon}
    \affiliation{Department of Physics, University of Basel, 4056 Basel, Switzerland} 
\author{Rafael S.~Eggli}
    \affiliation{Department of Physics, University of Basel, 4056 Basel, Switzerland}    
\author{Taras Patlatiuk}
    \affiliation{Department of Physics, University of Basel, 4056 Basel, Switzerland}

\author{Nathan Korda}
    \affiliation{Mind Foundry Ltd, Summertown, Oxford OX2 7DD, United Kingdom}  
\author{Dominik Zumb\"uhl}
    \affiliation{Department of Physics, University of Basel, 4056 Basel, Switzerland}    
\author{Natalia Ares}
    \thanks{natalia.ares@eng.ox.ac.uk}
    \affiliation{Department of Engineering Science, University of Oxford, Oxford OX1 3PJ, United Kingdom}

\begin{abstract}

Spanning over two decades, the study of qubits in semiconductors for quantum computing has yielded significant breakthroughs \cite{loss1998quantum, petta2005coherent, koppens2005control, xue2022quantum, noiri2022fast, mkadzik2022precision}. However, 
the development of large-scale semiconductor quantum circuits is still limited by challenges in efficiently tuning and operating these circuits. Identifying optimal operating conditions for these qubits is complex, involving the exploration of vast parameter spaces \cite{neyens2023probing}. This presents a real `needle in the haystack' problem, which, until now, has resisted complete automation due to device variability and fabrication imperfections \cite{zwolak2023colloquium, moon2020machine}.
In this study, we present the first fully autonomous tuning of a semiconductor qubit, from a grounded device to Rabi oscillations, a clear indication of successful qubit operation. We demonstrate this automation, achieved without human intervention, in a Ge/Si core/shell nanowire device. Our approach integrates deep learning, Bayesian optimization, and computer vision techniques. We expect this automation algorithm to apply to a wide range of semiconductor qubit devices, allowing for statistical studies of qubit quality metrics. As a demonstration of the potential of full automation, we characterise how the Rabi frequency and $g$-factor depend on barrier gate voltages for one of the qubits found by the algorithm. Twenty years after the initial demonstrations of spin qubit operation, this significant advancement is poised to finally catalyze the operation of large, previously unexplored quantum circuits.

\end{abstract}

\date{\today{}}
\maketitle




Recent advances underscore the potential of qubits in semiconductors for universal quantum computation~\cite{veldhorst2015two, burkard2023semiconductor, vandersypen2017interfacing, hensgens2017quantum, xue2022quantum, noiri2022fast, mkadzik2022precision, zhang2023universal}. These include the achievement of two-qubit gates showcasing fidelities that surpass thresholds essential for fault-tolerant computing~\cite{noiri2022shuttling, xue2022quantum, mills2022two}, and hot qubits that address the bottleneck of millikelvin refrigeration \cite{vandersypen2017interfacing, petit2020universal, camenzind2022hole}. Strides in wafer-scale manufacturing of these devices~\cite{neyens2023probing, zwerver2022qubits} and their efficient testing at cryogenic temperatures~\cite{de2023compact, thomas2023rapid} further highlight the rapid progress in this field.
Still, semiconductor quantum circuits are limited to at most six qubits \cite{philips2022universal} in one device. This stands in stark contrast to the potential afforded by modern semiconductor fabrication techniques, which could support the integration of millions of qubits.

 One of the reasons for this contrast is that a long-standing challenge remains: the intricate tuning required to reach and maintain qubit operation. 
Previous works have introduced diverse approaches for automating single stages of this process, such as defining double quantum dot (DQD) confinement potentials \cite{darulova2020autonomous, moon2020machine, severin2021cross, van2022all}, navigating to specific charge transitions \cite{kalantre2019machine, zwolak2020autotuning, zwolak2021ray, nguyen2021deep, baart2016computer, czischek2021miniaturizing, durrer2020automated, lapointe2020algorithm}, fine-tuning of charge transport features~\cite{vanesbroeck2020quantum} or the inter-dot tunnel couplings~\cite{teske2019machine, vandiepen2018automated}, as well as device characterisation \cite{chatterjee2022autonomous, craig2021bridging, schuff2023identifying}. 
Offering glimpses of the potential of machine learning for full qubit tuning automation, these works left the challenge unaddressed.

Here, we present a fully autonomous tuning process, able to encode a qubit without the need for human intervention. The process of going from a fully de-energized device to the observation of Rabi oscillations, a definitive indicator of qubit functionality, 
usually takes human experts weeks, or even months, to complete. Our algorithm, deployed on a DQD device, 
can complete the tuning process within three days.  

Our success in moving away from the manual tuning of semiconductor qubits marks a paradigm shift in quantum device scalability. Key to this success is the algorithm's ability to navigate through various stages of the tuning process, efficiently handling challenges and making accurate decisions. 
Our findings, underpinned by deep learning, Bayesian optimisation, and computer vision techniques, would finally 
allow for the operation and characterisation of complex, large-scale semiconductor qubit circuits. 

\begin{figure*}[ht]
\centering
  \includegraphics[width=1\linewidth]{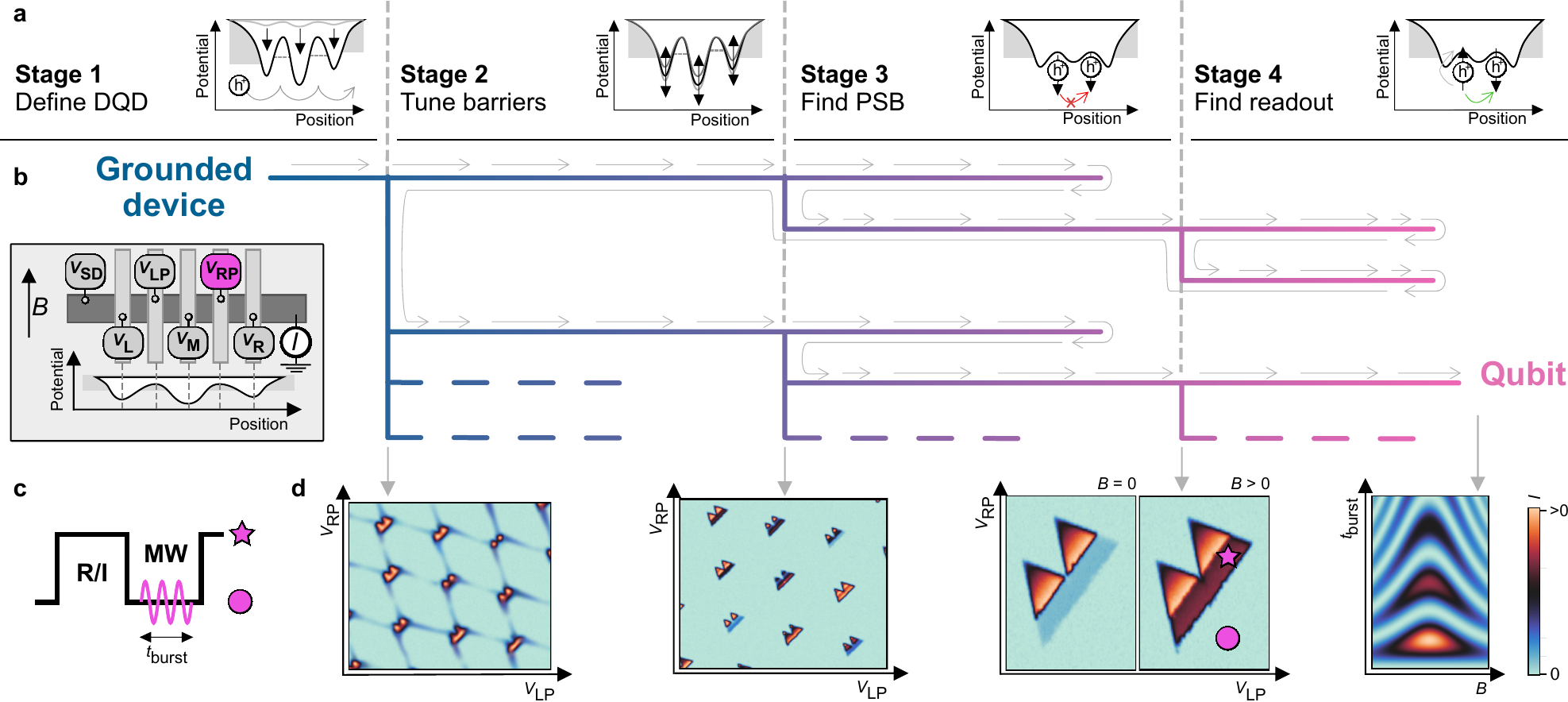}

  \justifiedcaption{ \textbf{Algorithm overview.} \textbf{a,} There are four stages that the algorithm needs to successfully navigate to reach qubit operations. The goal of each stage is illustrated with a confinement potential diagram. \textbf{b,} Each stage can either be successful and produce candidates (leading to one or more branches), or unsuccessful (leading to a backtracking to the closest stage that still has candidates). The search is therefore conducted in a tree structure. Some branches of the tree might be left unexplored. This is indicated by dashed lines. Inset: Illustration of the device. Five different voltages $V_i$ can be applied to a linear confinement. \textbf{c,} A pulsing scheme can be applied with a microwave burst of length $t_\text{burst}$. The fast line is connected to plunger gate RP. \textbf{d,} Simulated measurement illustrations that mark the successful transition between stages. Starting from a stability diagram with mere DQD features (far left), the algorithm tunes parameters until promising bias triangles (middle left), triangles exhibiting PSB (middle right) and finally Rabi oscillations (far right) are obtained. The extremal points of the pulse scheme are indicated as a star and circle in the middle right illustration.}
  \label{fig:algo_flow}
\end{figure*}

\section*{Device architecture and readout technique}

We consider a common layout for a DQD device (inset of Fig.~\ref{fig:algo_flow}b). We use a Ge/Si core/shell nanowire device in which holes are confined in depletion mode \cite{carballido2023qubit}. The electrical potential is set by a number of gate electrodes. Two plunger gate electrodes predominantly shift the electrochemical potential in the left and right dots with voltages $V_\text{LP}$ and $V_\text{RP}$. The rest of the gate electrodes primarily control the barriers between the DQD and the leads as well as the inter-dot coupling. 
One of the plunger gates is connected to a high-frequency line via a bias-tee, allowing for the application of voltage pulses and microwave bursts. 

The device can be probed by applying a bias voltage $V_\text{SD}$ to the source lead and recording the current $I_\text{SD}$ at the drain lead. 
The algorithm navigates to a DQD occupation that exhibits Pauli spin blockade (PSB) for spin-to-charge conversion. To achieve this, the DQD does not need to be depleted to the single hole regime. The charge occupation 
on each dot is estimated to be in the range of several dozens \cite{froning2021strong, Froning2018, ungerer2023charge, eggli2023cryogenic}. 

The algorithm uses a two-stage pulsing protocol \cite{camenzind2022hole, maurand2016cmos, hendrickx2020single, froning2021ultrafast} which is parameterised by a microwave (MW) pulse frequency and a duration $t_\text{burst}$(Fig.~\ref{fig:algo_flow}c). This protocol allows for qubit manipulation if the spin resonance conditions are met. Details on the pulsing scheme for coherent spin control and the device are described in \ref{methods} and by Carballido \textit{et al.}~\cite{carballido2023qubit}. 

\begin{figure*}
\centering
  \includegraphics[width=1\linewidth]{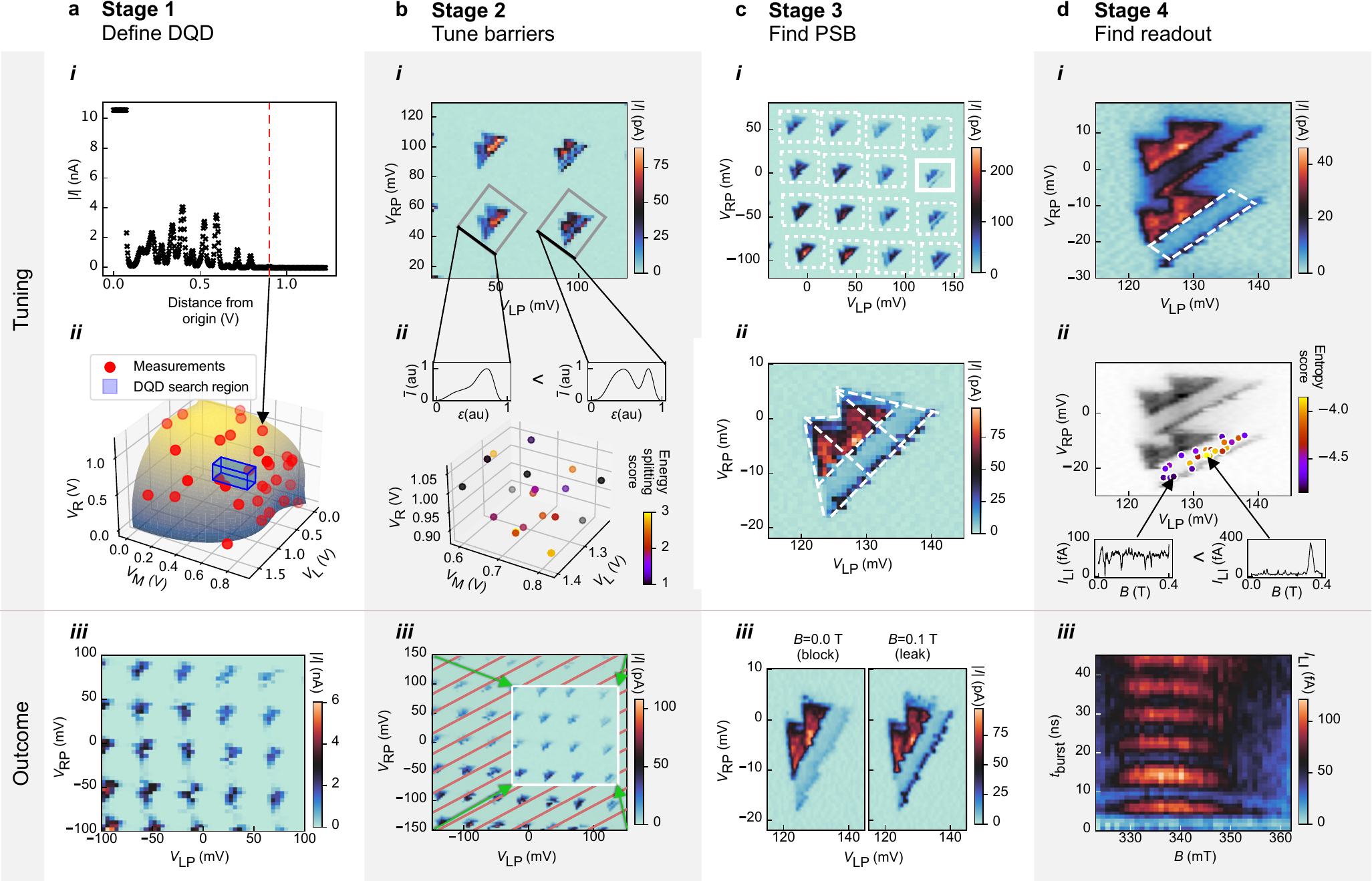}

  \justifiedcaption{ \textbf{Summary of each stage.} 
  \textbf{a,} Stage 1: Definition of a DQD potential. 
  \textbf{a-\textit{i}} Current measurements along random directions in gate voltage space to determine the points at which conducting and non-conducting regions meet, so-called pinch-off points.
\textbf{a-\textit{ii}} Gaussian processes model the hypersurface after collecting sufficient pinch-off points.
\textbf{a-\textit{iii}} A neural network confirms the presence of a DQD analysing the aquired stability diagrams.
  \textbf{b,} Stage 2: Optimisation of bias triangle features.
  \textbf{b-\textit{i}} Analysis of stability diagrams, segregating individual transitions and averaging them along segments orthogonal to the detuning axes.
  \textbf{b-\textit{ii}} Distribution of optimisation scores and averages of bias triangles along segments orthogonal to their detuning axes. We aim to increase the singlet-triplet energy splitting via a proxy score that measures the dip of current between the baselines and the rest of the triangles.
  \textbf{b-\textit{iii}} Identification of plunger voltage windows unaffected by charge switches aided by neural networks.
  \textbf{c,} Stage 3: Finding PSB
  \textbf{c-\textit{i}} Initial low-resolution, wide-range detection of PSB using neural networks.
  \textbf{c-\textit{ii}} Detailed high-resolution scans employing segmentation algorithms.
  \textbf{c-\textit{iii}} Bias triangles that show signatures of PSB and can be used to optimise readout.
  \textbf{d,} Stage 4: Readout spot identification within a promising transition
  \textbf{d-\textit{i}} Acquisition of stability diagrams pulsing gate RP to locate the readout region (indicated by a white dashed box).
  \textbf{d-\textit{ii}} Entropy-based scoring of magnetic field traces within the readout region, optimized through Bayesian methods.
  \textbf{d-\textit{iii}} Rabi oscillations for different magnetic fields around the resonance condition to confirm qubit operation. The measurements marked with $I_\text{LI}$ were amplified with a lock-in amplifier, see \ref{methods} for details.
  }
  \label{fig:algo_illustration}
\end{figure*}

\section*{The algorithm}

The autonomous tuning algorithm is structured into four main stages. Starting from a completely de-energized device, i.e.~with all gate voltages set to 0, the first two stages define the DQD potential by tuning the inter-dot barrier and the reservoir coupling. The third stage narrows the search space by looking for distinct signatures of PSB, an initialisation and readout requirement. The last stage fine tunes the plunger voltages and finds the frequency and duration of a microwave pulse needed to drive the qubit. The effect of each stage on the DQD confinement potential is illustrated in Fig.~\ref{fig:algo_flow}a. Measurement illustrations exemplifying those taken by the algorithm are shown in Fig.~\ref{fig:algo_flow}d.

As a result of the algorithm design a search tree emerges, as shown in Fig.~\ref{fig:algo_flow}b. Once a stage is successfully completed, a list of candidates is generated. A candidate consists of all information needed for the next stage to investigate it, usually containing locations or ranges of gate voltages, or information on the suspected $g$-factor and Rabi frequency $f_\text{Rabi}$. The candidates are ordered by a dedicated score in each stage and a single candidate is passed on to the next stage. 
If a stage is unsuccessful, the algorithm backtracks to the previous stage and investigates the next candidate in that stage's list of candidates. This process dynamically creates a search tree. If a different branch has proven to lead to a qubit, some branches of the tree may be left unexplored. These are indicated by dashed lines in the tree in Fig.~\ref{fig:algo_flow}. 

 We describe each stage in this section. Details on the stage structures, substages, and composition of candidates for each stage can be found in \ref{methods} and the Supplementary Material.


\subsection*{Stage 1: Define DQD}

The first stage identifies the gate voltage settings that define the DQD confinement potential. It determines a lower and upper limit for each barrier gate voltage, which is used in subsequent stages.

Building upon the methodologies of Moon \textit{et al}.~\cite{moon2020machine} and Severin \textit{et al}.~\cite{severin2021cross}, a hypersurface model is created to distinguish between conducting and non-conducting regions within the three-dimensional barrier gate voltage space.
The algorithm takes current measurements along random directions within this space (Fig.~\ref{fig:algo_illustration} a-\textit{i}), and models the hypersurface with a Gaussian process, as depicted in Fig.~\ref{fig:algo_illustration} a-\textit{ii}. We expect a DQD potential forming near a corner of the hypersurface in the first octant. To pinpoint this corner, three specific current measurements are conducted using only one of the gate electrodes at a time. The resulting coordinates are then projected onto the model of the hypersurface, setting the lower bounds of the region where DQDs are likely to be found. The upper bounds of the region are given by the coordinates of the single gate pinch-off voltages, i.e.~when the current drops to a value that is indistinguishable from the noise floor. The resulting box is labelled as `DQD search region' in Fig.~\ref{fig:algo_illustration} a-\textit{ii}.

A methodical search within this box is conducted by the algorithm. The algorithm samples locations in the DQD search region and investigates them, starting from the point nearest to the projected corner and progressing to higher gate voltages. At each location a one-dimensional trace of the plunger gate voltages is taken and checked for Coulomb peaks, a signature for quantised charge transport and a first requirement for a DQD. 
If Coulomb peaks are found, a measurement varying both plunger gate voltages, a so-called stability diagram, is acquired and analyzed via a neural network for DQD characteristics; see Fig.~\ref{fig:algo_illustration} a-\textit{iii} for a stability diagram that shows the desired features. 
Successful identification of DQD features at a location in gate voltage space establishes it as a lower limit for the subsequent search.



\subsection*{Stage 2: Tune barriers}

The goal in the second stage is to adjust the tunnel barriers to enhance the 
singlet-triplet energy splitting, an important requirement for qubit operation. Additionally, this stage needs to avoid regions of gate voltage space with following characteristics: regions with high currents above a generous threshold; regions that are susceptible to charge switch noise; regions that show co-tunneling lines.


Within the gate voltage bounds established in the first stage, Bayesian optimisation is employed to search gate voltage combinations. The figure of merit for this optimisation is based on the degree to which it reduces current between the triangle baseline and excited states, compared to the current throughout the rest of the bias triangles. This acts as an easy-to-compute proxy score of the singlet-triplet energy splitting. After Bayesian optimisation suggests a voltage setting, a stability diagram is measured (Fig.~\ref{fig:algo_illustration} b-\textit{i}). We first use a segmentation algorithm \cite{kotzagiannidis2023bias} to find the outlines of individual pairs of bias triangles. The score is then computed using averages of the stability diagram along the common baseline, as illustrated in Fig.~\ref{fig:algo_illustration} b-\textit{ii}. A neural network identifies bias triangles that are impacted by charge switches. Charge switches distort the stability diagrams and make the area unsuitable for qubit operation; see the hatched area in Fig.~\ref{fig:algo_illustration}b-\textit{iii} or Extended Data Fig.~\ref{fig:switch_illustration} for examples. These bias triangles are excluded from the optimisation. The gate voltage regions explored by the optimisation are shown in Fig.~\ref{fig:algo_illustration} b-\textit{ii}.   

A significant challenge of this stage is the need for numerous stability diagrams, which are time-intensive to measure. To address this, we implement an adaptive, efficient measurement algorithm designed to specifically focus on gate voltage regions where bias triangles are present, see Supplementary Material for details. Employing this method cuts down the measurement time by approximately two-thirds. The optimisation is performed using these efficient measurements.

As a final step, stability diagrams without the efficient measurement algorithm are taken in the most promising regions. This is done to ensure there are no charge switches, because the previous step ranked each location by the highest score of a bias triangle at that location. Therefore, some stability diagrams may have regions affected by charge switches. Bounding boxes are then created in the plunger voltage space, encompassing primarily stable bias triangles with current below the previously mentioned threshold (Fig.~\ref{fig:algo_illustration} b-\textit{iii}). These triangles are each evaluated and scored, and the bounding boxes are ranked based on the highest score they contain.

The algorithm has up to this point only used a positive bias voltage. This stage proposes both positive and negative bias voltages for each candidate it creates. The gate voltage coordinates including the bias voltage are passed on to the subsequent stage. 


\subsection*{Stage 3: Find PSB}

In this stage, our algorithm searches for charge transitions exhibiting PSB, a necessary condition for qubit initialisation and readout in this setup. A candidate has to pass three different classifiers to be judged as exhibiting PSB. This is necessary to avoid false positives entering the time-intensive last stage.

We begin with low-resolution stability diagrams of bias triangle candidates, both with $B=0$ T and $B=0.1$ T. In these devices, PSB is expressesd as a suppressed baseline of the bias triangles in low magnetic field compared to high magnetic fields (Fig.~\ref{fig:algo_illustration} c-\textit{iii}).


We use a routine based on the auto-correlation of the stability diagram to pinpoint bias triangle locations, see \ref{methods} for details. 
The identified triangles are then analyzed using a neural network
~\cite{schuff2023identifying}(Fig.~\ref{fig:algo_illustration} c-\textit{i}). Subsequently, the algorithm measures a stability diagram to precisely delineate the bias triangle, followed by high-resolution stability diagrams with $B=0$ T and $B=0.1$~T.

The algorithm invokes a routine to segment the bias triangles \cite{kotzagiannidis2023bias}, illustrated in Fig.~\ref{fig:algo_illustration} c-\textit{ii}. A further PSB classification based on the segmentation is performed. The routine from \cite{kotzagiannidis2023bias} defines the direction in plunger voltage space that controls the detuning $\varepsilon$ of the dot energies, known as the detuning axis of the bias triangles. Scanning along this line with varying magnetic fields, we expect to observe a current drop at the baseline at zero magnetic field, which another classifier detects.
Upon meeting all criteria, the gate voltage coordinates are forwarded to the next stage.



\begin{figure*}[htbp]
\centering
  \includegraphics[width=1\linewidth]{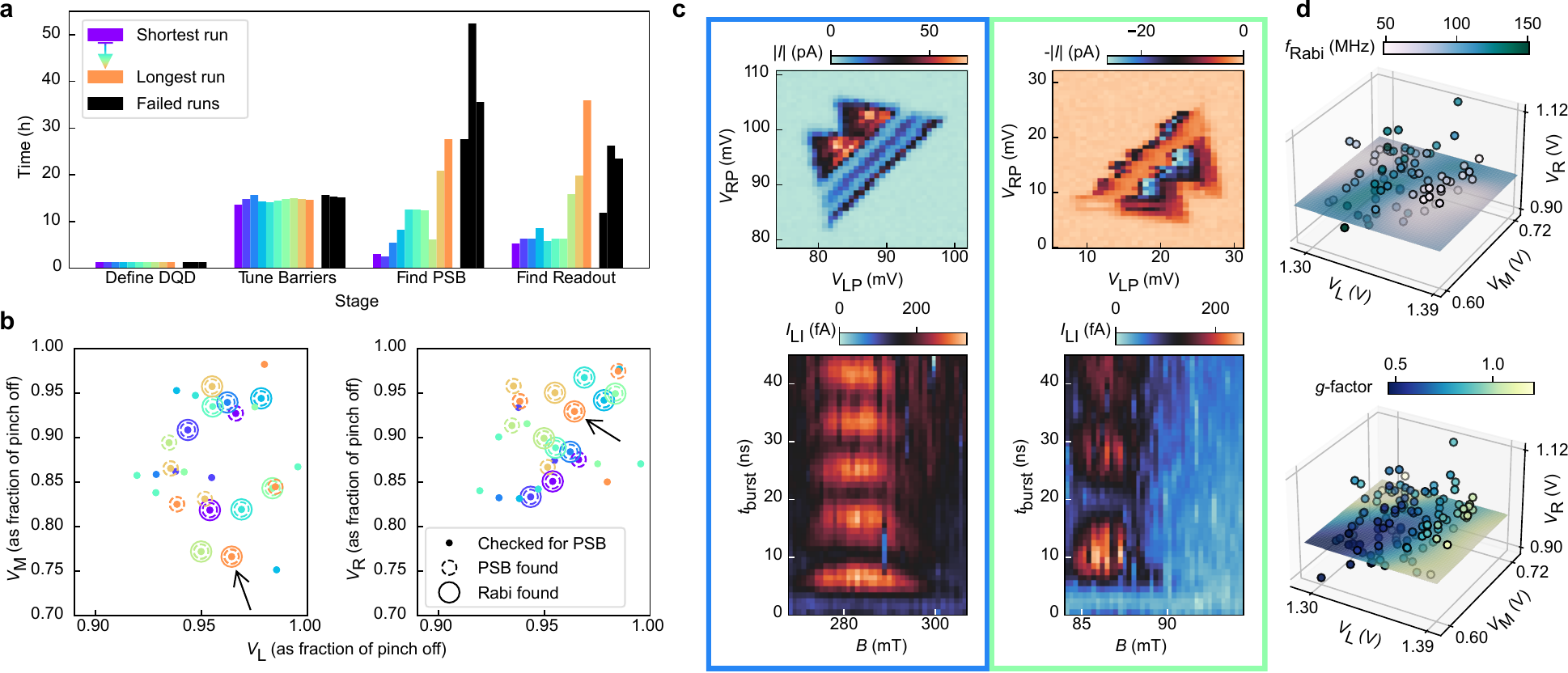}

  \justifiedcaption{ \textbf{Benchmarking.} \textbf{a,} The duration of each stage for ten successful runs is shown, sorted by the total time taken to tune a qubit in operation. In black: the three runs that did not lead to a qubit. \textbf{b,} Locations in the barrier voltage space (as a fraction of the individual pinch-off voltage) is shown. In some locations, no PSB was identified. The qubit locations are a subset of locations that exhibit PSB. \textbf{c,} Examples of transitions from two different runs, showing the associated bias triangles (upper row) and Rabi chevrons (lower row). \textbf{d,} Characterisation. Varying the barrier gate voltages, we can build a map of $f_\text{Rabi}$ (top row) and the $g$-factor (bottom row) of one of the found qubits (marked with arrows in b) using automated measurements. For illustrative purposes we show slices of a Gaussian Process model that was fitted to the data.}
  \label{fig:results}
\end{figure*}
\subsection*{Stage 4: Find readout}

The final stage of the process is dedicated to finding an operating point for qubit readout and manipulation. 
This stage not only identifies a suitable location in plunger voltage space but also determines the optimal driving frequency and duration of the pulse.

Based on the segmentation from the previous stage, a predicted readout gate voltage region within the bias triangle is defined (Fig.~\ref{fig:algo_illustration} d-\textit{i}). The algorithm optimizes across the four-dimensional space of the two plunger gate voltages, the driving frequencies and pulse durations. It samples a point within this space and then measures the current as a function of the magnetic field (Fig.~\ref{fig:algo_illustration} d-\textit{ii}). The optimisation's goal is to identify a current peak in these scans, indicative of the qubit's resonance condition (Fig.~\ref{fig:algo_illustration} d-\textit{ii}, lower right plot). This is achieved by evaluating the entropy of current traces; traces exhibiting a peak correspond to lower entropy values, see \ref{methods} for details. 


Bayesian optimisation then proposes potential candidates for further analysis. These candidates are filtered based on the presence of one or two peaks (corresponding to the number of qubits addressed), as determined by a simple peak-finding algorithm \cite{2020SciPy-NMeth}. Noisy measurements might also show peaks. Therefore, a follow-up step involves retaking the measurement to confirm the presence of a peak.

Once a candidate is verified, several measurements are taken by the algorithm to establish the qubit's operational functionality. These include a spectroscopy measurement, a Rabi chevron experiment (Fig.~\ref{fig:algo_illustration} d-\textit{iii}), and repeated high resolution Rabi oscillations at the Larmor frequency.

\subsection*{Hyperparameters}
Each stage requires a set of hyperparameters. They control various aspects of the measurements such as: resolution and safe gate voltage ranges for stability diagrams; aspects of the signal processing algorithms such as the required prominence of peaks; and steering parameters such as the number of candidates that a stage can suggest. The measurement aspects can be derived from some weak prior knowledge about device such as the magnitude of lever arms which informs the resolution of stability diagrams. We also assume a $g$-factor larger than 0.5 (limiting $B$ and $f_\text{MW}$) and a $f_\text{Rabi}$ between $30$ MHz and $250$ MHz (limiting $t_\text{burst}$). The requirements on the prior knowledge can be easily softened by widening the search space.

The set of hyperparameters influences the length of the runs and the way the algorithm manages trade-offs between exploration and exploitation. Regardless of the hyperparameters, the algorithm will always terminate once all candidates of each stage have been exhausted.

The choice of hyperparameter was made during development and  not optimised for total run time or efficiency. 
We provide a full list of all hyperparameters in the Supplementary Material.

\subsection*{Characterisation}

After a qubit has been found, we may choose to study the qubit further. Routines from Stages 3 and 4 enable the tracking of a known readout spot and the resonance condition in gate voltage space. This allows for the recording the dependency of qubit metrics as we change the confinement potential. We provide details of this characterisation algorithm in \ref{methods}.

\section*{Results}


We fixed the hyperparameters and gathered 13 runs.
Rabi oscillations were found in ten of those runs. 
In successful runs, the time spent in each stage varied (Fig.~\ref{fig:results}a).
total time required ranged from 22 to 80 hours, with a mean of 38 hours (median 34 hours) (Table \ref{tab:metrics}). Each stage relies on the exploration and accuracy of the previous stage. The variation in the time required in each stage gets progressively larger. 
Almost all time is spent on measuring the device, not on the decision algorithms. This is due to the measurement of current through the DQD which requires long integration times. A setup that allows for fast measurements via, e.g.~radiofrequency reflectometry could be tuned orders of magnitudes faster.

\begin{table}[!ht]
    \centering
    \begin{tabular}{r||l|l|l|l|l|l|l|l|l|l}
    \toprule
        Run \# & 1 & 2 & 3 & 4 & 5 & 6 & 7 & 8 & 9 & 10 \\ \midrule \midrule
        Time & 22.8 & 24.5 & 28.5 & 32.0 & 33.3 & 34.4 & 34.5 & 37.8 & 56.5 & 79.2\\
        (h)  & ~ & ~ & ~ & ~ & ~ & ~ & ~ & ~ & ~ & ~   \\ \midrule
        $f_\text{Rabi}$& 31 &	97&	109&	47&	51&	90&	56&	115&	49&	63 \\ 
        (MHz) & \tiny$\pm$1 & \tiny$\pm$2 & \tiny$\pm$1 & \tiny$\pm$1 & \tiny$\pm$1 & \tiny$\pm$1 & \tiny$\pm$1 & \tiny$\pm$3 & \tiny$\pm$1 & \tiny$\pm$1 \\ 
        $g$ & 1.52&	0.72&	0.70&	2.20&	2.74&	0.73&	2.31&	0.67&	2.12&	0.74 \\
        &   \tiny $\pm$0.07&  \tiny$\pm$0.05&  \tiny$\pm$0.05&  \tiny$\pm$0.12&  \tiny$\pm$0.22&  \tiny$\pm$0.04&  \tiny$\pm$0.14&  \tiny$\pm$0.08&  \tiny$\pm$0.14&  \tiny$\pm$0.03\\\midrule
        $V_\text{L}$ (V) & 1.34 & 1.34 & 1.34 & 1.37 & 1.36 & 1.35 & 1.35 & 1.32 & 1.35 & 1.38 \\ 
        $V_\text{M}$ (V) & 0.69 & 0.81 & 0.79 & 0.80 & 0.69 & 0.78 & 0.79 & 0.77 & 0.65 & 0.71 \\ 
        $V_\text{R}$ (V) & 0.91 & 1.02 & 0.95 & 1.01 & 1.04 & 0.94 & 0.95 & 0.89 & 1.00 & 1.02 \\ \bottomrule
    \end{tabular}
    \justifiedtablecaption{\textbf{Metrics of successful runs.} For each qubit found, we show the total time it took, the Rabi frequency $f_\text{Rabi}$, the $g$-factor, and the settings for the barrier gates $V_\text{L}$, $V_\text{M}$ and $V_\text{R}$. The errors of $f_\text{Rabi}$ are estimated from the fit uncertainty and the errors of $g$ are calculated from the width of the resonance peak.}
    \label{tab:metrics}
\end{table}

The three failed runs terminated after their exploration on a similar timescale to the longest successful runs. They took between 56.0 h and 94.9 h to complete and are shown in black in Fig.~\ref{fig:results}a. The runs found transitions that showed Pauli spin blockade but the algorithm was not able to find Rabi oscillations, possibly due to unfortunate settings of the tunnel barrier strengths in Stage 2. 

The qubit locations in gate voltage space in terms of the three barrier gate voltages are depicted in Fig.~\ref{fig:results}b. For comparability, we normalise each voltage by the voltage at which each barrier gate electrode pinches off the current individually. At each point, Stage 2 (Tune barriers) passed a candidate for further analysis (solid dots). In some cases, PSB was detected (dashed circles), passing Stage 3, and a subset of these also yielded a qubit (solid circles), successfully completing Stage 4.

Fig.~\ref{fig:results}c presents examples from two runs, showing the transitions found and Rabi chevron measurements. The discovery of qubits in both bias directions evidences the algorithm's adaptability and its non-specificity to certain transitions. Both Rabi chevrons were obtained using the same given driving power and driving frequency, but varied in magnetic fields and Rabi frequencies, highlighting the algorithm's generalization capability. All depicted measurements were autonomously executed by the algorithm, accounting for the non-centered chevron measurements.


Analyzing the locations in gate voltage space where qubits were found provides insight into the device physics (Table \ref{tab:metrics}). By fitting a convex hull around the qubit locations in the barrier gate voltage space, we can estimate the volume of the region where qubits can be found. For this device, the volume of the convex hull is approximately $3.5\times10^{-4}$V$^3$, translating to a fraction of the safe ranges [$(2 \text{V})^3$] of about $4\times10^{-5}$. The space is further restricted by the plunger voltage location, which is a box of roughly $(10 \text{mV})^2$. Given a search space of $(300 \text{mV})^2$, this brings down the size of the volume to around $2\times10^{-7}$ as a fraction of the 5-dimensional gate voltage space. That is roughly equivalent to a needle in a $(2\text{m})^{3}$ haystack.

Once a qubit has been found, the algorithm allows for extensive characterisation. We can study 
$f_\text{Rabi}$ and the $g$-factor as a function of the barrier gate voltages (Fig.~\ref{fig:results}d). The resulting maps gives insights into qubit properties and can be extended to measure, e.g.~the Hahn-echo coherence time.

 


\section*{Conclusions}
We have demonstrated fully automated tuning of spin qubits, progressing from a de-energised device to qubit control. Our algorithm autonomously achieved Rabi oscillations in 10 out of 13 trials. 
Most tuning processes concluded within three days, with the primary speed constraint being the integration times required to perform DC transport measurements, which could be replaced by fast readout alternatives. 
Maps of the $g$-factor and $f_\text{Rabi}$ serve as evidence for the potential of this approach for high-throughput qubit characterisation.




The methodology is versatile and can be adapted for use with similar quantum devices, such as silicon FinFETs. The modular design of the algorithm allows for rapid adaptation to other architectures. For example, devices that use charge sensors and reflectometry for measurements would only require different signal-processing algorithms.

We anticipate that the mass tuning and characterization of qubits, facilitated by our fully autonomous tuning algorithm, will establish a productive feedback loop between measurement and fabrication processes. Wafer-scale, high-throughput characterisation of quantum devices, already feasible in early tuning stages, can mitigate device variability. This, in turn, improves the tuning process, bolstered by expanding datasets.

The first successful experiments on quantum computing with semiconductors were conducted nearly twenty years ago. 
We have now confirmed the feasibility of fully automatic tuning of a semiconductor qubit. This breakthrough would allow us to move forward by leveraging the ability to mass characterise qubits to advance quantum circuit scalability in semiconductors.
\bibliography{reference}

\newpage

\section*{Methods}


Here we provide details of the experimental set up, and algorithmic choices made in this study. We begin by describing the device used for our experiments. We then follow with a discussion of the intelligent computational approaches used in the algorithm. Finally, details of each stage are discussed. An overview of all the stages of the algorithm is given in Table \ref{tab:all-stages}.

Hyperparameters that are not explicitly given here can be found in the Supplementary Material.

\makeatletter\def\@currentlabel{Methods}\makeatother
\label{methods}

\section{The device, measurement apparatus and pulse sequence}
The device consists of a Ge/Si core/shell nanowire lying on top of nine bottom gates measured in a variable temperature insert (VTI) in a liquid helium bath with the sample mounted below the 1-K pot (base temperature 1.5 K). By applying positive voltages to the first five bottom gates from the left, an intrinsic hole gas inside the nanowire is depleted to form a hole DQD. An SEM image of a device similar to the one used here is shown in Fig.\ref{fig:device}. The device, measurement apparatus and pulse sequence are the same as used by Carballido \textit{et al.}~\cite{carballido2023qubit} and are described in detail in their work.

To amplify the measurements that rely on a microwave pulse, we applied a pulse-modulation by a lock-in amplifier at 87.777 Hz. The measurements therefore have an in-phase and out-of-phase component. We apply principal component analysis (PCA) to these measurements and project each measurement onto the principal axis. We further offset the measurements such that they are strictly positive. Measurements that are obtained this way are marked with $I_\text{LI}$, as opposed to currents that were measured conventionally which are marked with $|I|$.

\section{Techniques used in the algorithm}
Useful automation of tuning from a de-energised device to identification of Rabi oscillations requires an algorithm that can adapt to different data capture regimes, and be transferred to other, similar devices. To achieve this, we have made extensive use of intelligent, adaptive and data-driven subroutines. Nowadays, there is a plethora of such techniques to choose from, and in order to choose the right technique to apply to each stage of the algorithm, we considered:
\begin{itemize}
    \item the need for expert labelled training data, which was not always possible or realistic to source;
    \item the need of being efficient both in the total number of measurements taken during a stage, and in the resources needed for computing decisions about which measurements to take;
    \item the minimal accuracy needed in each stage for the whole algorithm to be able to achieve its overall goal of identifying operating parameters for a qubit.
\end{itemize}
To address the above considerations across the many stages of the algorithm, a non-exhaustive list of techniques we have found useful to employ includes Gaussian process (GP) inference, convolutional neural networks (CNNs), unsupervised computer vision (CV) and computational geometry techniques, and Bayesian optimisation (BO). We now briefly introduce these, highlighting their strengths and weaknesses.

GPs are a popular form of non-parametric Bayesian inference \cite{Rasmussen2005GaussianPF}. They can be thought of as a method for doing principled Bayesian inference over a space of functions. GPs can be tailored to any specific domain or problem by making a choice of the so-called kernel (or covariance) function, a part of this model which describes prior knowledge about the possible space of functions in which inference is to occur. This choice can allow practitioners to encode important domain knowledge before capturing any data, such as specifying knowledge of periodicities, symmetries or the expected degree of smoothness of the underlying process that is being observed. This constitutes the main strength of GP modelling, often enabling highly data-efficient inference. However, both model fitting and model prediction can be computationally intensive, typically growing cubically \cite{Rasmussen2005GaussianPF} in the number of observed data points.

Over the past two decades, CNN architectures have proved to be go-to models for solving computer vision tasks. Their strengths lie in their adaptability across different computer vision tasks, their robustness in the face of unknown noise, and their computational efficiency at training time. Their weakness lies in always requiring substantial amounts of training data. In this work, we use some standard architectures, such as ResNet \cite{He2015DeepRL} as tools for extracting properties from or making assessments of stability diagrams. Where we have applied them, training data has been either generated by a sufficiently good simulator, or gathered from this or similar devices and then labelled by an expert.

Often, the use of CNNs is neither required nor appropriate for the particular computer vision task at hand. In particular, it has been of crucial importance to a number of stages through the algorithm to be able to automatically locate and segment bias triangles within stability diagrams using a coordinate-wise approach. To achieve this we have employed a number of unsupervised computer vision techniques that can mitigate noise in, and localize features of the geometric figures present in stability diagrams. Whereas CNNs would require large amounts of labelled training data to achieve this result, the requirements can be met by computer vision techniques that need no training data, and require only a few hyperparameter choices to be made. All together we have called this a bias triangle segmentation framework, and it is described in \cite{kotzagiannidis2023bias}, where the specific application to PSB detection is also detailed.

Bayesian optimisation \cite{Shahriari2016TakingTH} is a general, iterative approach to black-box function optimisation. At each iteration, it constructs a surrogate model of the function being approximated using the data already gathered, and uses this surrogate model to efficiently compute the next most informative location from which to sample the unknown function. In order to apply this technique one must specify a score function to optimise and a parameter space over which the search for an optimum is conducted. The choice of the surrogate model is also influential in the accuracy, efficiency and reliability achieved using this method. In this work we have made consistent use of GP surrogate models using a Matern 5/2 kernel.

\section{Stages of the algorithm}

\subsection{Define DQD}
\subsubsection{Hypersurface building}

As a first step, the algorithm determines a current that it considers to be pinched off by ramping to the high end of the safe ranges. We take repeated current measurements there to characterise the noise floor. From the noise floor we compute a pinch off current. 

Next, we sample several points within the safe ranges using a Sobol sequence for quasi-random locations. The points are used to define rays from the origin that are then investigated for pinch off. To avoid overloading the current amplifier, we search for pinch off from the origin towards the upper end of the safety ranges with a low bias voltage. Once pinch off is found, we retrace with a higher bias voltage to confirm the exact pinch off location. 

Finally, we use this data to construct the hypersurface model as outlined in the main text.

%



\subsubsection{Double dot detection}
The previous stage defines a region within which we can look for a DQD potential. We sample quasi-random points via a Sobol sequence for investigation. For each point, we vary both plunger voltages simultaneously and measure the current, following the method described by Moon \textit{et al.}~\cite{moon2020machine}. We use the random forest classifier developed by Severin \textit{et al.}~\cite{severin2021cross} to check for the presence of Coulomb peaks. If Coulomb peaks are found, we then measure a stability diagram. This diagram is analyzed with a neural network to detect features of the DQD. The neural network was trained on data from a variety of devices, mainly from the data gathered by Moon \textit{et al.}~\cite{moon2020machine} and Severin \textit{et al.}~\cite{severin2021cross}, and additional data from a nanowire device different from the one used in this work. In total, there were 4,611 stability diagrams of which 726 showed double dot features.


\subsection{Tune barriers}
\subsubsection{Barrier optimisation}
Upon formation of separated pairs of bias triangles, we perform coordinate-wise segmentation and polygon fitting (see Sect.~\ref{app:segmentation} 
for more details, \cite{kotzagiannidis2023bias}) in order to facilitate their tracking and feature extraction throughout the remainder of the tuning pipeline.

The segmentation and shape extraction enable the assessment of the current intensity difference between the base line and gap formed between the base and first excited state line. By quantifying this intensity difference in a base separation score, we can employ a Bayesian optimisation framework which seeks to maximize the intensity difference, as a promising indicator for detecting viable candidates for PSB.

Each pair of bias triangles should present a base well separated from the main body. The wider the separation, the better the protection of the ground state and the more likely it is that we are in the presence of a qubit. The separation score is computed by averaging the current along the detuning axis, see the two one-dimensional traces in Fig.~\ref{fig:algo_illustration} b-ii, and computing the ratio of the intensity between the peaks and the lowest point in the valley between them. In the case of multiple triangles, the highest separation is used as score.

Certain potential landscapes can lead to situations in which charge configurations are affected by charge switch noise.  
Since these potentials are unlikely to be used as a qubit and the resulting bias triangles can skew our base separation scoring, we excluded them by leveraging a neural network classifier. This classifier was trained to distinguish between normal bias triangles and ones that are affected by charge switch noise, as illustrated in Fig.~\ref{fig:switch_illustration}. The training dataset for this classifier was obtained as follows: initially, potential bias triangles were identified using our segmentation routine. Subsequently, we hand-labeled 2,302 of these (1,539 samples showed no switch noise) to create a robust training set. The classifier itself was then obtained by fine-tuning a ResNet-based architecture with this dataset. 

Once the voltage space has been explored through BO we have a clear understanding of the landscape, as in Fig.~\ref{fig:algo_illustration} b-ii. The measurements in this optimisation were obtained using an efficient measurement algorithm, see Supplementary Materials for details. We sample the most promising regions again without the efficient measurement algorithm and analyse them as described in the next section. 


\subsubsection{Plunger window selection}
Given a sampled stability diagram containing bias triangles, the aim is to select the region that contains as many bias triangles with scores as high as possible, no areas with current that is too high, and as few switches as possible as depicted in Fig.~\ref{fig:algo_illustration} b-iii. High current bias triangles are unlikely to be able to be used for qubit operations. This is a heuristic and we set a conservative threshold of 200 pA. To ease the downstream steps, the region should be a rectangular window. Through an iterative approach, starting from the smallest bounding boxes containing each single pair of bias triangles, larger windows are constructed by merging the existing ones in case they satisfy the conditions about switch absence and low currents. For the switches absence we used a soft constraint, allowing for bias triangles affected by switch noise in case their total area was less than 25\% of the area covered by all triangles in the window. The algorithm complexity scales exponentially with respect to the number of triangles and some heuristics have been leveraged to reduce substantially complexity and therefore execution time. In particular, at each iteration, only the top 100 bounding boxes by the number of contained triangles without switches were kept, to ensure a manageable upper limit on the number of possible merges. The routine halts when no further merges are possible.  
Once the plunger windows have been selected they are ranked by highest separation score. 

\subsection{Find PSB}
\subsubsection{Wide shot PSB detection}
To identify each bias triangle's location, we first leverage the fact that they sit on a honeycomb or skewed rectangle pattern. We use autocorrelation on the stability diagram to identify this pattern. The largest two peaks in the autocorrelation help us establish a vector that spans this pattern of skewed rectangles. To fix the pattern in place, we employ a blob detection algorithm, using the first blob it identifies. This helps us accurately overlay the skewed rectangle pattern and estimate the locations of the bias triangles (Fig.~\ref{fig:autocorrelation}).

Next, we extract these bias triangles using the identified locations, with side lengths informed by the pattern dimensions. These extracted bias triangles are then input into a neural network for further analysis. We used autonomously gathered data that was taken during the initial development phase. In total, we used 626 pairs of bias triangles taken from 70 stability diagrams. 55 of the pairs of bias triangles showed PSB. For more detailed information on this procedure, see Schuff \textit{et al.}~\cite{schuff2023identifying}.




\subsubsection{Re-centering and high resolution PSB detection}
\label{app:segmentation}
In an effort to filter the previously detected 
candidates for PSB and eliminate false positives, a second set of higher resolution measurements is performed. For that purpose, a dedicated low resolution stability diagram of the candidate bias triangles is taken and used to update the plunger voltage extent based on the detected contours, effectively performing a re-centering. With the updated voltage extent, high resolution stability diagrams with $B = 0 $ T and with $B=0.1$ T are taken.\\
\\
A second sub-stage of PSB-classification is applied through a segmentation-based detection and feature extraction framework, which facilitates the coordinate-wise quantification of geometric and physical properties of bias triangles \cite{kotzagiannidis2023bias}.
In particular, given the high-resolution stability diagram with $B=0.1$ T, this framework fits minimum-edge polygons to the detected contours of bias triangle pairs by utilizing a relaxed extension of the Ramer-Douglas-Peucker algorithm \cite{kotzagiannidis2023bias}.
Once the segmented shape mask is identified, further geometric properties such as the base and excited state lines can be automatically extracted solely based on the prior knowledge of the bias voltage sign, which predicts the direction in which bias triangles point.\\
\\
For the identification of PSB, an analytical classifier based on the above framework was devised \cite{kotzagiannidis2023bias}. PSB expresses itself as a suppressed base line of the bias triangles at $B=0$ T. At $B>0$ T, there is a leakage current at the base line. The routine extracts the segment enclosing the base and a prominent excited state line on the stability diagram with leakage current ($B=0.1$ T). Subsequently, the average pixel intensity of the segment normalised by the intensity of the entire pair of bias triangles is computed. By superimposing the detected segment on the scan with blocked current ($B=0$ T), normalised intensity values are compared and, if their difference exceeds a specified threshold, the charge transition is identified as positive for PSB.\\
\\
Based on the segmented bias triangle shape mask, further geometrical properties can be automatically extracted, which enables the tuning of bias triangle features in stage 2.
\\
The detuning axis, utilized for the Danon gap measurement, is automatically extracted by identifying the bias triangles base midpoints and tips and computing the lines between them.

\subsubsection{Danon gap check}
\label{app:danon_gap_check}
As a further filter for possible candidates, we check to magnetic field dependence of the leakage current at the base of the bias triangles in a different way. As a function of the applied magnetic field $B$, we expect the leakage current to be minimal at $B=0$ T and higher away from this point. We call this the Danon gap \cite{danon2009pauli}. A current measurement of the detuning line while varying the magnetic field gives us a two-dimensional input (Fig.~\ref{fig:danon_gap}), which we can analyse as follows: Ignoring the noise signal, the current is roughly constant along the magnetic field axis, whereas the detuning line axis is information-rich: away from the Danon gap there are two local extrema (to one side the noise floor outside the bias triangles, to the other side the gap due to singlet-triplet energy splitting) whilst the Danon gap region is characterised by a monotonic behaviour, with roughly a constant value. 
\\
To detect the presence of the Danon gap, the current $I$ is firstly processed with a Gaussian filter, to smooth out the noise, and then the absolute slopes along the detuning line axis are integrated $g(B):= \sum_{\varepsilon}{\left|\frac{\partial \tilde{I}(\varepsilon,B)}{\partial \varepsilon}\right|}$, where the derivative is the discrete derivative along the detuning line axis. The function $g$ is minimised in the areas where the smoothed signal $\tilde{I}$ shows a constant value. We show the normalised function $\bar{g} = \frac{g}{|\varepsilon|}$ in Fig.~\ref{fig:danon_gap}. To detect the presence of the Danon gap from $g$ two tunable hyperparameters are used, validating the depth and width of the basin of the global minimum of $g$: in case the basin is not prominent enough there is no Danon gap. As the last check, the location of the minima has to be in proximity of zero magnetic field.
\\

\subsection{Find readout}
\subsubsection{Tracking \& entropy optimisation} 
In subsequent steps, we apply a pulse sequence to the right plunger electrode. As it is a two-stage pulse, the bias triangles will have a `shadow' in the stability diagram. We need to identify the original bias triangles and find a suitable region where we can expect to find qubit readout. 
In light of resulting shape distortions and further degrading effects to the measurement quality, we opt for template matching as opposed to performing re-segmentation for bias triangle tracking in order to ensure robustness.\\
\\
The relative direction in which the shadow bias triangles appear with respect to the original one is known in advance due to the applied pulse shape. This is incorporated into the shape matching approach as the cardinal direction to uniquely identify and track the triangles.
\\
We perform shape matching by comparing the edge map of a stability diagram prior to pulsing, functioning as the template, to the edge map of a subsequent stability diagram with pulsing, functioning as the source for current information. Further, we extract the segmented shape mask from the template.
The method slides the template over the source edge map, thereby comparing the template with individual patches of the stability diagram with pulsing, and returns a result matrix (of the same size as the source) whose individual entries quantify the similarity with the template patch. The employed similarity metric is the normalized correlation coefficient and the patch with maximum
correlation is selected. Once the appropriate patch has been identified, the initial segmentation mask of the stability diagram without pulsing is mapped to the stability diagram with pulsing and used for subsequent processing. 
\\
In order to identify the optimal readout spot, we extract the segment enclosing the base and prominent excited state lines on the obtained segmented mask of bias triangles with pulsing, and perform Bayesian optimisation of a readout quality score over four parameters: the constrained two-dimensional plunger gate voltage space, frequency of the driving pulse $f_\text{MW}$ and burst time $t_\text{burst}$.
\\
Optimal readout candidates are those which meet the resonance conditions of the qubit. If they are met, there is a leakage current that we record using the lock-in amplifier. For a given burst time $t_\text{burst}$ (relating to the Rabi frequency $f_\text{Rabi}$) and a given frequency of the driving pulse $f_\text{MW}$ (relating to the $g$-factor), the leakage is characterized as a peak in leakage current for a certain magnetic field $B$. Thus, for the readout optimisation, we use measure the current with varying magnetic fields. Instead of applying PCA, as explained above, we use the L2 norm of the in-phase and out-of-phase components to retrieve a one-dimensional trace $l(B)$. This guarantees peaks to be higher than the background, as opposed to processing with PCA, which can lead to dips rather than peaks.

To quantify the sharpness of these peaks
, we developed a score based on the Shannon entropy $H = -\sum_B \left[l(B) \log(l(B)\right]$ of the trace. For the calculation of the entropy of the score, we first subtract the median and then clip values at zero. This particular pre-processing turns the trace into something more akin to a distribution and enhances the robustness of our score, making it less susceptible to potential noise disturbances in the trace. This method results in a smooth score landscape suitable for Bayesian optimisation.

\subsubsection{Resonance confirmation}
This verification step acts as a final filter and the last stages are all executed once a candidate passes this filter. The previous stage sends a candidate with a suspected resonance condition. The stage re-measures the leakage current as a function of the magnetic field. If the resonance condition was truly found, a peak should appear at the same magnetic field again. Should a peak with a specified prominence at this magnetic field within a set margin of error be detected, the resonance condition is considered confirmed and all downstream measurements are executed.
We note that a noisy candidate might pass this stage. Repeating the verification step can reduce such occurrences.



\subsubsection{Qubit measurements}


%

Once a resonance condition is found, we vary the magnetic field and the burst duration. The characteristic Rabi chevron can be analysed by considering the frequency spectrum for each magnetic field. The frequency should have a minimum at the magnetic field that meets the resonance condition of the qubit. The amplitude should also be maximal there due to decoherence for off-resonant driving.
We can therefore simply look for the maximum amplitude in the Fourier transformed Rabi chevron (Fig.~\ref{fig:chevron_analysis}). This information will give us the precise resonance conditions for the last step, repeated measurements of Rabi oscillations on resonance.

\section{Characterisation}

The maps in Fig.~\ref{fig:results} were generated using automated measurements. Initially, upon identifying a qubit, we record its readout spot, $g$-factor, and $f_\text{Rabi}$. We then alter the confinement potential by slightly adjusting the barrier gate voltages. This adjustment may shift the bias triangles, consequently moving the readout spot, $g$-factor, and $f_\text{Rabi}$. Our method involves tracing these transitions to locate the readout spot in its new position. At this new location, we conduct an EDSR check scan. Any changes in the peak's location inform us about variations in the $g$-factor. Furthermore, measuring Rabi oscillations at this point helps update our understanding of $f_\text{Rabi}$.

As we progressively deviate from the initial measurement point, we utilize our closest prior qubit data to infer the properties at the new location. This step is crucial as a shift in the $g$-factor necessitates modifying the magnetic field range, while a change in $f_\text{Rabi}$ requires adjusting the $t_\text{burst}$ duration for the EDSR check to accurately detect resonances.





\section*{Acknowledgements}

This work was supported by the European Research Council (grant agreement 948932), the Royal Society (URF-R1-191150), Innovate UK project AutoQT (grant number 1004359), NCCR SPIN of the Swiss NSF, Swiss Nanoscience Institute (SNI), the EU H2020 European Microkelvin Platform EMP (grant no. 824109), FET TOPSQUAD (grant no. 862046), and FETFLAG QLSI (grant no. 951852).
J.S.~acknowledges financial support from the EPSRC (grant number R72976/CN001).

We want to thank Andrew Briggs, Georgios Katsaros, Dominic Lennon and Leon Camenzind for helpful discussions.

\section*{Author Contributions}


J.S.~developed the algorithm, in particular the modular framework and the overall algorithm flow.
J.S.~performed the experiments and analysed the results with M.J.C.'s guidance and help from T.P.~and R.S.E.~in D.M.Z.'s lab. M.J.C.,~S.S.~and P.C.K.~fabricated the device. J.S., M.J.C., J.C.C., M.K., M.C., J.R., N.K., D.C., B.v.S.~and B.S.~developed the routines for the various stages of the algorithm, including analysing and interpreting data, making algorithm design choices, and writing the software used to run and record the experiments.
J.C.C., M.K., M.C., J.R., and N.K. made substantial independent contributions to the Methods and Supplementary Materials sections.
J.S.~wrote the manuscript with inputs from all authors.
N.A.~conceived of the project.

\section*{Data and Code Availability}
The data and code will be made available upon final publication.




\newpage

\appendix

\clearpage

\section*{Extended Data}\label{app:...}

\begin{figure}[!htbp]
\centering
  \includegraphics[width=.66\linewidth]{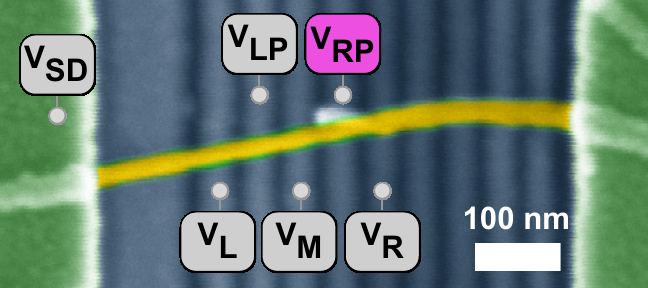}

  \justifiedcaption{ \textbf{Device.}  False color micrograph of a nanowire device similar to the one used in this work. The Ge/Si nanowire is colored in yellow.}
  \label{fig:device}
\end{figure}

\FloatBarrier
\begin{figure}[!htbp]
\centering
  \includegraphics[width=1\linewidth]{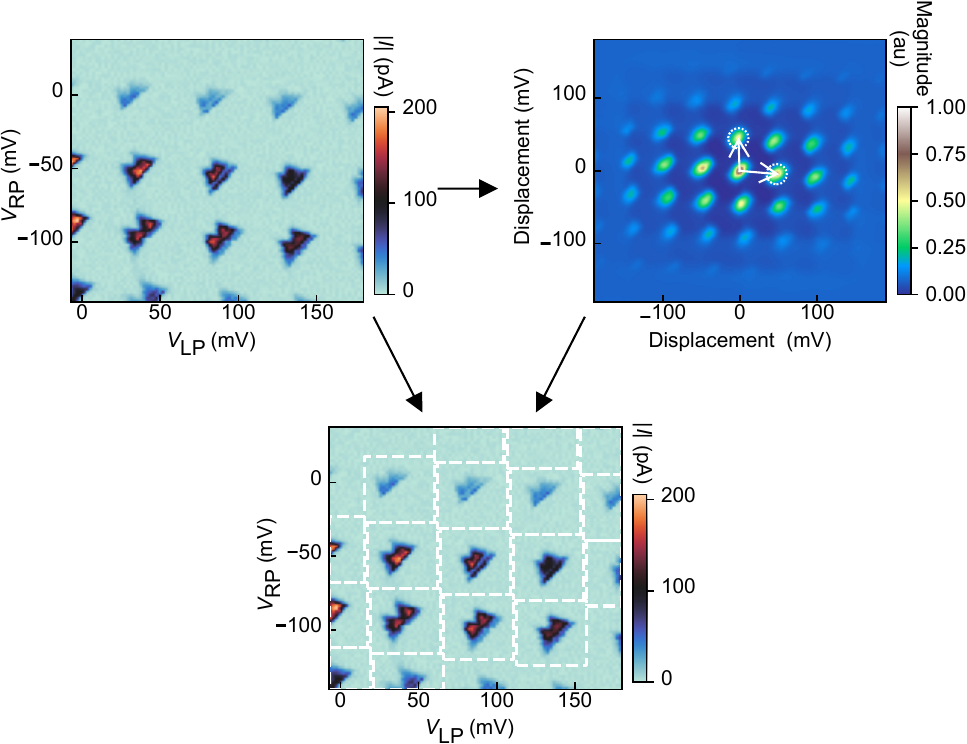}

  \justifiedcaption{ \textbf{Bias triangle locations via auto-correlation.} A stability diagram (top left) is processed using its auto-correlation (top right). Within the auto-correlation, we can find the highest values to span a grid of skewed rectangles. The extracted information is used to locate bias triangles in the original stability diagram (bottom).
  }
  \label{fig:autocorrelation}
\end{figure}
\FloatBarrier

\begin{figure}[htbp]
\centering
  \includegraphics[width=1\linewidth]{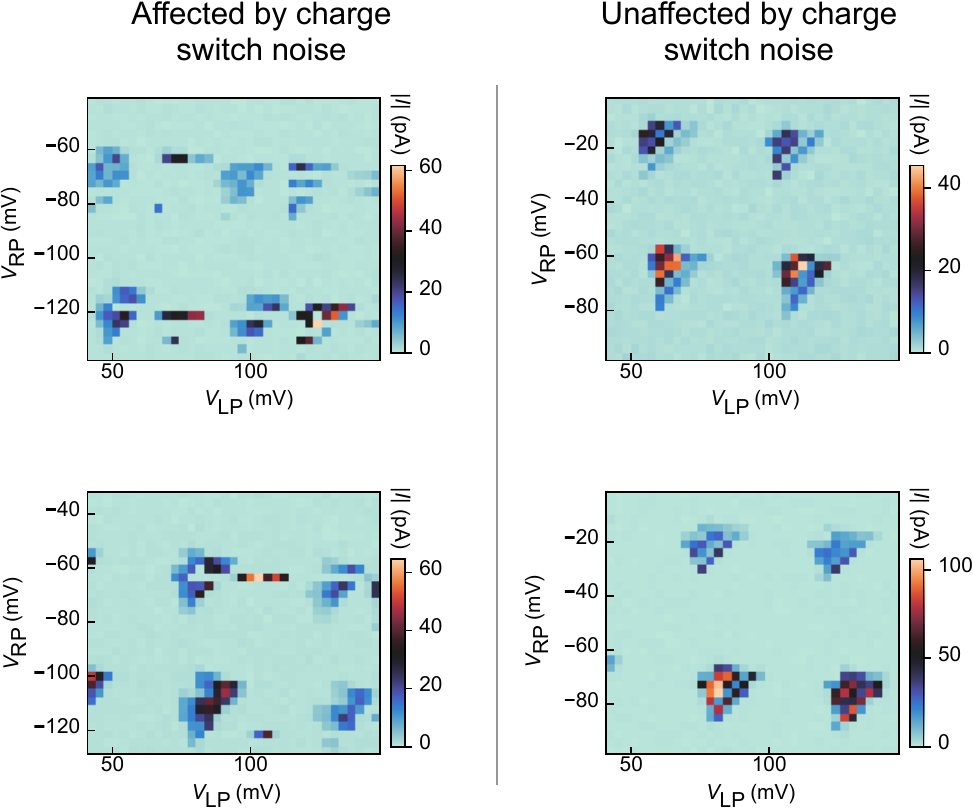}

  \justifiedcaption{ \textbf{Effect of charge switch noise.} Stability diagrams can be affected by charge switches. Their effect can be seen in the left column. For comparison, the right column is unaffected by these charge switches.}
  \label{fig:switch_illustration}
\end{figure}

\FloatBarrier

\begin{figure}[!htbp]
\centering
  \includegraphics[width=1\linewidth]{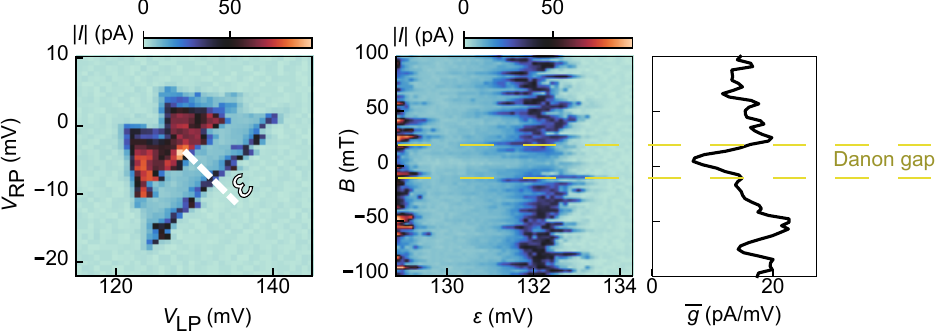}

  \justifiedcaption{ \textbf{Danon gap check.} As final check to confirm the presence of PSB, the algorithm takes measurements of the detuning line $\varepsilon$ (marked in the left panel) as a function of the magnetic field $B$ (middle). The measurement is analysed, as described in \ref{app:danon_gap_check}, by first computing the averaged sum of absolute derivatives $\bar{g}$ (right). If a dip is present in $\bar{g}$ close to $B=0$ T, PSB is confirmed.}
  \label{fig:danon_gap}
\end{figure}

\begin{figure}[!htbp]
\centering
  \includegraphics[width=1\linewidth]{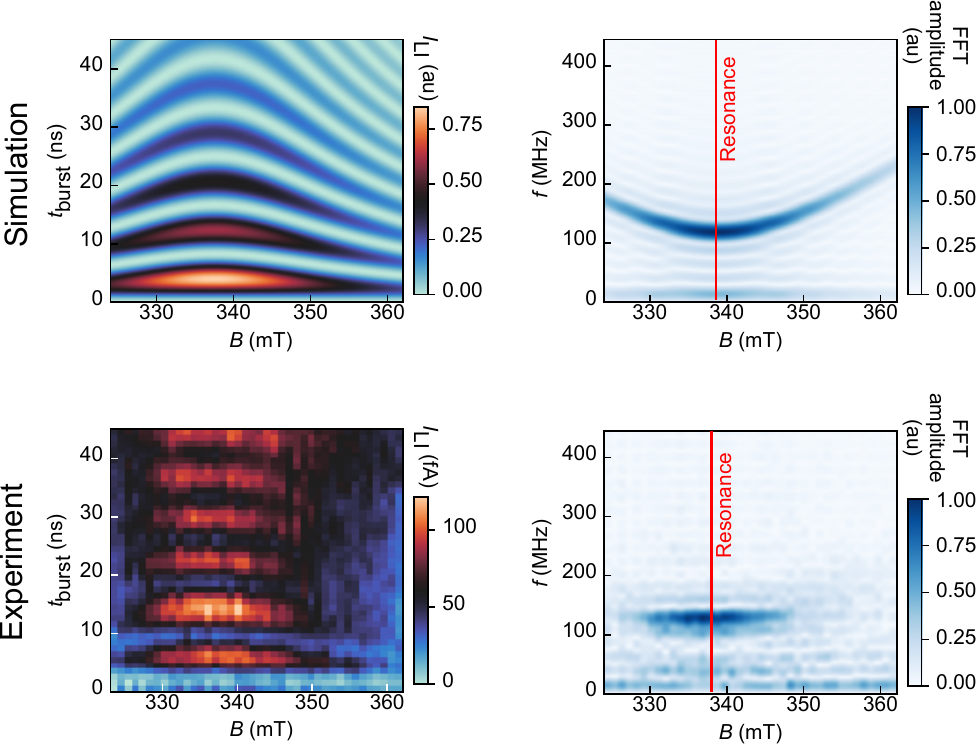}

  \justifiedcaption{ \textbf{Rabi chevron analysis.} A qubit that is driven close to the resonance condition exhibits so-called Rabi chevrons, see left column for examples (top: simulation, bottom: experimental data). To find the resonance frequency, we analyse the frequency spectrum for each magnetic field (which controls the detuning from resonance). The frequencies have a maximum amplitude and a minimal point at the resonance condition.}
  \label{fig:chevron_analysis}
\end{figure}

\begin{table*}[htbp]

  \centering
    \begin{tabularx}{\textwidth}{l||X|X|X}
            Stage                    & Description                                                                        & Techniques used                                                             & Candidate information                                             \\
\toprule
\toprule
\textbf{Stage 1:} Define DQD \\
\textbf{a,} Hypersurface building  & Building a model of the surface that separates pinch-off from conducting  & Gaussian processes & Upper and lower bounds of barrier voltages  \\
\textbf{b,} Double dot detection  & Identifying Coulomb peaks and double dot signatures & Random forests and neural networks & Corrected lower bound of barrier voltages \\
                  
    \midrule
                
\textbf{Stage 2:} Tune Barriers\\
\textbf{a,} Barrier optimisation& Search over the barrier voltage space for ideal settings & Bayesian optimisation, computer vision, neural networks& Promising barrier voltage location \\
\textbf{b,} Plunger window selection & Determine a window for plunger voltages & Computer vision, neural networks& Barrier voltage location and wide plunger voltage range \\

    \midrule
            
\textbf{Stage 3:}   Find PSB  \\
\textbf{a,} Wide shot PSB detection& Identify locations of transitions with PSB & Neural networks, computer vision & Narrow plunger voltage range \\
\textbf{b,} Re-centering & Get precise plunger voltage window & Computer vision & Corrected narrow plunger voltage range \\
\textbf{c,} High res.~PSB detection & Confirm PSB with high resolution measurement & Computer vision & Narrow plunger voltage range, link to high res.~measurement, detuning line definition\\
\textbf{d,} Danon gap check & Confirm PSB by measuring detuning line as a function of magnetic field & Computer vision & Narrow plunger voltage range, link to high res.~measurement \\

    \midrule
    
\textbf{Stage 4:} Find Readout\\
\textbf{a,} Entropy optimisation & Find readout spot, $g$-factor and Rabi frequency by optimisation of an entropy score & Bayesian optimisation, computer vision & Precise plunger voltage locations, magnetic field, drive frequency and burst time \\
\textbf{b,} Resonance confirmation & Confirm resonance from previous stage/filter out noise& Peak finding & Passed on from previous \\
\textbf{c,} Spectroscopy& Measurement of current while varying $t_\text{burst}$ and $f_\text{MW}$ for documentation& -& Passed on from previous\\
\textbf{d,} Rabi chevron& Measuring Rabi oscillations close to the resonance condition & Frequency analysis& Corrected magnetic field for resonance condition\\
\textbf{e,} Rabi oscillations& Take repeated Rabi oscillations on resonance & - & - \\                                                 
\bottomrule
\end{tabularx}
\justifiedtablecaption{\textbf{List of all stages.} We list all stages and sub-stages with a short description, a rough overview of what techniques were used and what the information is passed downstream for a candidate from each stage. Besides information on the parameters that are directly needed to operate a qubit, the stages also pass down meta-information that other stages might need to use, e.g.~the high resolution stability diagram of Stage 3c is needed in Stage 4a for template matching.}
\label{tab:all-stages}
\end{table*}%



\end{document}


\title{Supplementary Information: Fully autonomous tuning of a spin qubit}

\author{Jonas Schuff}
    \affiliation{Department of Materials, University of Oxford, Oxford OX1 3PH, United Kingdom}
\author{Miguel J. Carballido}
    \affiliation{Department of Physics, University of Basel, 4056 Basel, Switzerland}    
\author{Madeleine Kotzagiannidis}
    \affiliation{Mind Foundry Ltd, Summertown, Oxford OX2 7DD, United Kingdom}
\author{Juan Carlos Calvo}
    \affiliation{Mind Foundry Ltd, Summertown, Oxford OX2 7DD, United Kingdom}
\author{Marco Caselli}
    \affiliation{Mind Foundry Ltd, Summertown, Oxford OX2 7DD, United Kingdom}
\author{Jacob Rawling}
    \affiliation{Mind Foundry Ltd, Summertown, Oxford OX2 7DD, United Kingdom}
\author{David L. Craig}
    \affiliation{Department of Materials, University of Oxford, Oxford OX1 3PH, United Kingdom}
\author{Barnaby van Straaten}
    \affiliation{Department of Materials, University of Oxford, Oxford OX1 3PH, United Kingdom}
\author{Brandon Severin}
    \affiliation{Department of Materials, University of Oxford, Oxford OX1 3PH, United Kingdom}
\author{Federico Fedele}
    \affiliation{Department of Engineering Science, University of Oxford, Oxford OX1 3PJ, United Kingdom}
\author{Simon Svab}
    \affiliation{Department of Physics, University of Basel, 4056 Basel, Switzerland} 
\author{Pierre Chevalier Kwon}
    \affiliation{Department of Physics, University of Basel, 4056 Basel, Switzerland} 
\author{Rafael S.~Eggli}
    \affiliation{Department of Physics, University of Basel, 4056 Basel, Switzerland}  
\author{Taras Patlatiuk}
    \affiliation{Department of Physics, University of Basel, 4056 Basel, Switzerland}

\author{Nathan Korda}
    \affiliation{Mind Foundry Ltd, Summertown, Oxford OX2 7DD, United Kingdom}  
\author{Dominik Zumb\"uhl}
    \affiliation{Department of Physics, University of Basel, 4056 Basel, Switzerland}    
\author{Natalia Ares}
    \affiliation{Department of Engineering Science, University of Oxford, Oxford OX1 3PJ, United Kingdom}

\date{\today}
\maketitle

\onecolumngrid

\tableofcontents

\newpage

\section{Qubit measurements of all successful runs}

\begin{figure}[!htbp]
\centering
  \includegraphics[width=1\linewidth]{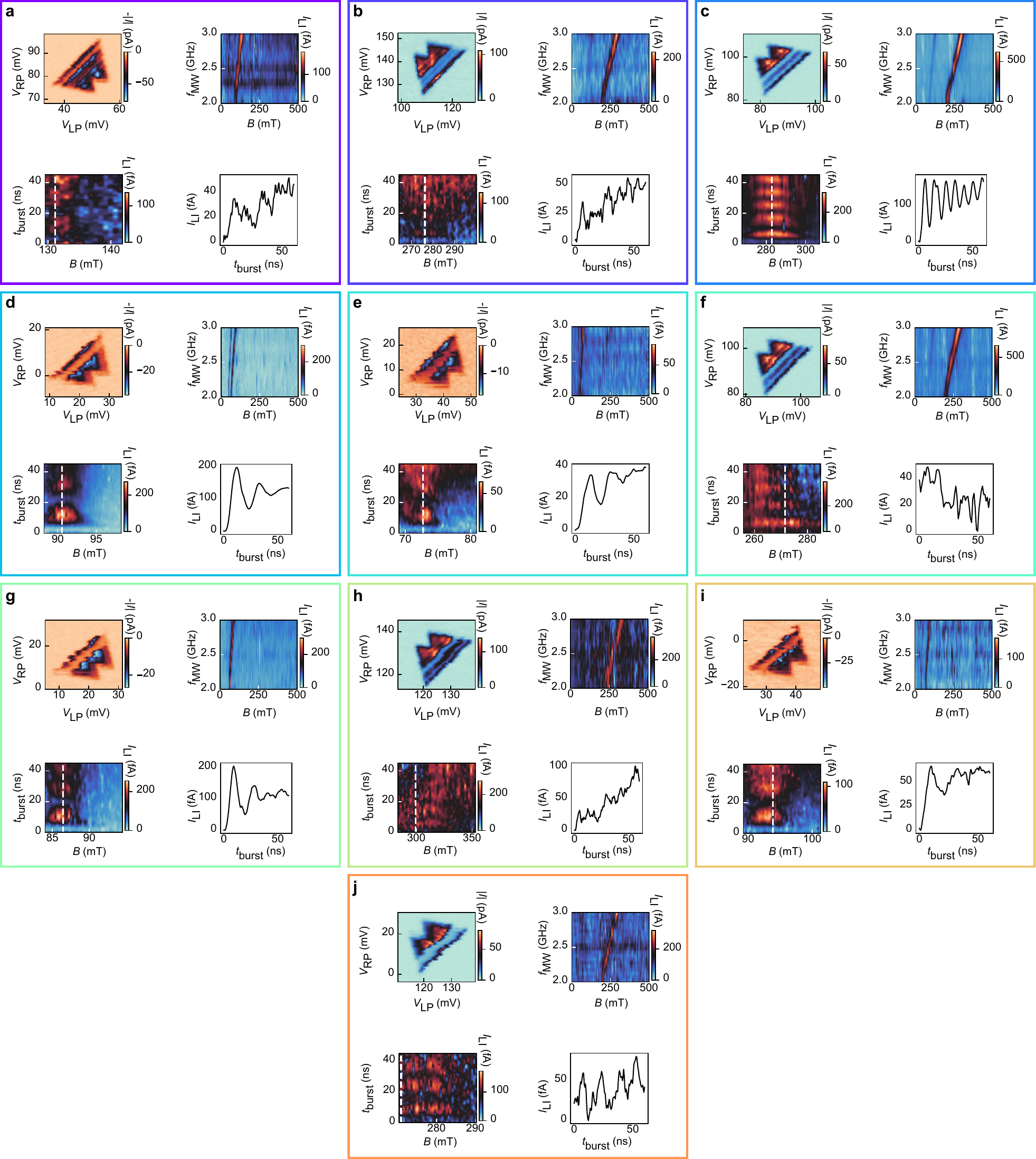}

  \caption{ \textbf{Qubit measurements of all successful runs.} The panels \textbf{a,-j,} are ordered by the total run time of the algorithm for each qubit respectively. Each panel includes four current measurements: the pair of bias triangles (upper left), spectroscopy measurement, varying magnetic field and driving frequency (upper right), Rabi chevron pattern, varying magnetic field and burst duration (lower left), and averaged Rabi oscillations (lower right) taken at the dashed lines in the Rabi chevron measurement. All measurements were performed autonomously. The Rabi chevron measurement does not have a dedicated re-centering stage, accounting for the off-centered measurements. The spectroscopy measurements were purely taken for documentation and always with the same ranges; these measurements did not inform any other part of the algorithm. Some measurements for panels \textbf{d}, \textbf{e}, and \textbf{f} were taken again using automated measurements after the initial runs finished because a setting of the lock-in amplifier led to slight measurement artifacts.}
  \label{fig:all_qubits}
\end{figure}

We show measurements that confirm that we found a qubit in the ten successful runs, ordered by total run time as in the main text, see Fig.~\ref{fig:all_qubits}. The measurements were all taken autonomously. We show the associated pair of bias triangles (upper left in each panel), a measurement varying the magnetic field and the driving frequency (upper right), a Rabi chevron measurement where we vary the magnetic field and the burst duration (lower left), and an averaged measurement of Rabi oscillations (lower right) at the magnetic field indicated with dashed lines in the Rabi chevron measurement.

The diversity in the plunger voltage settings, the magnetic field settings, and the Rabi frequencies showcase the versatility of our algorithm.

\clearpage

\section{Hyperparameters}
\label{supp:hyperparameters}

List of all hyperparameters and comment on what they do. We do not report on hyperparameters that are irrelevant to the functionality of the algorithm, such as how often measurements are plotted to a documentation file.

\subsection{Define DQD}
\subsubsection{Hypersurface building}
\begin{outline}
    \1 Steering parameters: 
    \2 \verb|number_of_rays| = 32 // Number of rays used to build hypersurface.
    \item \verb|n_noise_floor| = 100 // Number of measurements of the noise floor.
    \1 Measurement parameters:
    \2 \verb|lower_bounds| = $[0,0,0]$ // Defines lower bound of area in which the hypersurface model is built, in V.
    \item \verb|upper_bounds| = $[1.8,1.8,1.8]$ // Defines upper bound of area in which the hypersurface model is built, in V.
    \item \verb|bias_low| = 0.0007 // Low bias voltage used to find pinch off starting from the conducting region, in V.
    \item \verb|bias_high| = 0.005 // Bias voltage used to in subsequent stages. Used to confirm pinch off going from non-conducting to conducting region, in V.
    \item \verb|d_r| = 0.003 // Step length of pinch off search, in V.
    \item \verb|len_after_pinchoff| = 0.250 // Length after last point above threshold before pinch off is considered true (past last Coulomb peaks), in V.
    \item \verb|max_dist| = 2 // Additional safe range how far the ray is maximally ramped, in V.
    \1 Analysis parameters:
    \2  \verb|threshold_as_multiple_of_noise_high| = 100 // Used to define the noise threshold based on the noise floor measurement

\end{outline}    
    
\subsubsection{Double dot detection}
\begin{outline}

    \1 Measurement parameters:
    
    \2 \verb|magnetic_field| = 0.1 // At which magnetic field the measurements are taken, in T.

    \2 \verb|plunger_location| = $[0 ,0]$ // Center of the plunger gate voltages, in V.

    \2 One-dimensional scan (to detect Coulomb peaks):
        \3 \verb|window_right_plunger| = 0.1 // Defines side length of window in which measurements us taken, in V.
        \3 \verb|window_left_plunger| = 0.1 // Defines side length of window in which measurements us taken, in V.

        \3 \verb|n_px| = 128 // Number of points to be taken.
        \3 \verb|wait_time| = 0.051 // Delay after setting parameter before measurement is performed, in s.

    \2 Two-dimensional scan (to detect DQD features):
        \3 \verb|window_right_plunger| = 0.2 // As above.
        \3 \verb|window_left_plunger| = 0.2 // As above.

        \3\verb|n_px_rp| = 48 // Number of points to be taken for right plunger axis.
        \3 \verb|n_px_lp| = 48 // Number of points to be taken for left plunger axis.

        \item \verb|wait_time_slow_axis| = 0.5 // Delay after setting parameter before measurement is performed, in s.
        \item \verb|wait_time_fast_axis| = 0.051 // Delay after setting parameter before measurement is performed, in s.
    
    \1 Analysis parameters:
    
\2 \verb|max_distance_between_locations| = 0.1 // Used to determine number of points sampled within DQD search region. Sets maximal distance between each sampled point, in V.

    \2 \verb|path_to_nn| = \verb|local_path| // Path to weights of neural network used for DQD feature detection.

\end{outline}

\subsection{Tune barriers}
\subsubsection{Entropy optimisation}
\begin{outline}
        \1 Steering parameters: 
        \2 \verb|seeding| = 15 // Parameter for Bayesian optimisation that informs exploration period.
        \item \verb|n_required_results| = 30 // Number of stability diagrams to be taken.
        
    \1 Measurement parameters:
        \2 \verb|rp_start| = $-0.15$ // Starting point of measurement for right plunger, in V.
        \item \verb|rp_end| = 0.15 // End point of measurement for right plunger, in V.
        \item \verb|lp_start| = $-0.15$ // Starting point of measurement for left plunger, in V.
        \item \verb|lp_end| = 0.15 // End point of measurement for left plunger, in V.
        \item \verb|n_points_plungers| = 100 // Number of points in each dimension.

\end{outline}

\subsubsection{Plunger window detection}
\begin{outline}
        \1 Steering parameters: 
    
    \2  \verb|number_of_full_scans_threshold| = 10 // Maximum number of full scans (i.e., without the efficient measurement algorithm) to be taken.
    \item \verb|number_of_candidates| = 10 // Maximum number of candidates Stage 2 can suggest in each bias direction, i.e., 10 can lead to up 20 candidates.
    \item \verb|bias_directions| = [\verb|positive_bias|, \verb|negative_bias|] // Candidates are built in those bias directions. 
\end{outline}

\subsection{Find PSB}

\subsubsection{Wide shot PSB detection}
\begin{outline}
        \1 Steering parameters: 

    \2 \verb|max_number_candidates| = 5 // Maximum number of candidates this sub-stage can create.

    \1 Measurement parameters:

    \2 \verb|low_magnetic_field|= 0.0 // Magnetic field at which the stability diagram with blocked current shall be taken, in T.
    \item \verb|high_magnetic_field|= 0.1 // Magnetic field at which the stability diagram with leakage current shall be taken, in T.
    \item \verb|resolution| = 0.002 // Resolution of stability diagram in each axis, in V.
    \item \verb|padding| = 0.03 // Padding added to the plunger window suggestion from previous stage. Needed to have a slight margin around bias triangles, in V.

    \1 Analysis parameters:
    
    \2 \verb|psb_threshold| = 0.5 // Threshold for PSB detection. Neural network returns a value between 0 and 1 for each pair of bias triangles.
    \item \verb|folder_path_to_nn| = \verb|local_path| // Path to neural network model that predicts signatures of PSB from low resolution measurements.
    \item \verb|offset_px| = 10 // Parameter used in the location detection via auto-correlation. The highest peak will always be in the center, so peaks within a certain distance (given in pixel here) from the center are disregarded.

\end{outline}

\subsubsection{Re-centering}
\begin{outline}
    \1 Measurement parameters:

    \2 \verb|magnetic_field|= 0.0 // Magnetic field at which the stability diagram shall be taken, in T.

    \item \verb|resolution| = 0.002 // Resolution of stability diagram in each axis, in V.
    \item \verb|wait_time_slow_axis| = 0.5 // As above.
    \item \verb|wait_time_fast_axis| = 0.051 // As above.
    \1 Analysis parameters (all related to routine from Kotzagiannidis \textit{et al.}~\cite{kotzagiannidis2023bias}):
    \2 \verb|segmentation_upscaling_res| = 2 // Image is upscaled by this factor to improve segmentation.
    \item \verb|relative_min_area| = 0.01 // Computes the \verb|min_area| as a fraction of the total area. \verb|min_area| sets a threshold for the minimum area of contour to be detected to avoid outliers.
    \item \verb|denoising| = true // Apply Gaussian smoothing.
    \item \verb|allow_MET| = false // Determines whether the 'Minimal enclosing triangle' technique is used or not. If true facilitates enclosing triangle shape approximation for disconnected contours.
    \item \verb|thr_method| = 'triangle' // Thresholding method for contour detection.

\end{outline}

\subsubsection{High resolution PSB detection}

\begin{outline}
    \1 Measurement parameters:

    \2 \verb|low_magnetic_field|= 0.0 // As above.
    \item \verb|high_magnetic_field|= 0.1 // As above.

    \item \verb|resolution| = 0.00075 // As above.
    \item \verb|wait_time_slow_axis| = 0.5 // As above.
    \item \verb|wait_time_fast_axis| = 0.051 // As above.

    \item \verb|padding| = 0.005 // As above.

    \1 Analysis parameters:
    
    \2 \verb|slope_tol| = 0.4 // Tolerance for deviation in absolute value between slopes of detected lines.
    \item \verb|int_tol| = 0.05 // Tolerance for PSB metric (absolute value difference between normalized segment intensities).
    \item \verb|seg_tol| = 0.05 // Gives percentage of image length as threshold for segments that are too small.
    \item \verb|median| = false // If true, selects the median of detected lines (ordered by y-intercept); false by default, so that the line with largest y-intercept (outmost) is selected.
    \item \verb|segmentation_upscaling_res| = 2 // As above.
    \item \verb|relative_min_area| = 0.01 // As above.
    \item \verb|denoising| = true // As above.
    \item \verb|allow_MET| = false // As above.
    \item \verb|thr_method| = 'triangle' // As above.

\end{outline}

\subsubsection{Danon gap check}

\begin{outline}
    \1 Measurement parameters:
    
    \2 \verb|magnetic_field_min| = - 0.1 // Start of magnetic field, in T.
    \item \verb|magnetic_field_max| = 0.1 // End of magnetic field, in T.
    \item \verb|resolution_magnet| = 0.003 // Resolution of magnetic field, in T.

    \item \verb|resolution_detuning| = 0.0001 // Resolution of detuning line axis, in V.

    \item \verb|detuning_base_offset| = 0.002 // We add this to the detuning line measurement to include the full base as the segmentation algorithm can lead to detuning line definitions that end on the base line, therefore missing valuable information.

    \item \verb|extra_wait_time_slow_axis| = 0.5 // The slow axis (magnetic field) is delayed by the time needed for the magnet to ramp one position, plus this given time.
    \item \verb|wait_time_fast_axis| = 0.077 // As above.

    \1 Analysis parameters:

    \2 \verb|segmentation_upscaling_res| =2 // As above.
    \item \verb|min_area| = 3 // As above.
    \item \verb|thr_method| = 'triangle' // As above.
    \item \verb|allow_MET| = false // As above.
    \item \verb|padding_factor| = 1 // As above.
    \item \verb|minimum_det_line_length_ratio| = 0.33 // The detuning line is determined via the segmentation algorithm. It also determines a cutoff within the triangles so that the algorithm only takes measurements at the base line of the triangle. If the detuning line that is determined is less than \verb|minimum_det_line_length_ratio| of the full detuning line (from base line to the tip of the triangles), we extend the detuning line definition to avoid detuning lines definitions that are too short.
    \item \verb|peak_offset_tolerance| = 0.025 // The gap can be at most offset from 0T by this much and still be accepted as a true gap, in T.

    \item \verb|sigma| = 1 // For the gap detection, Gaussian smoothing factor.
    \item \verb|field_gap_size| = 0.002 // For the gap detection, parameter that controls maximal gap width.
    \item \verb|relative_depth| = 1.0 // For the gap detection, parameter that controls maximal gap depth.

\end{outline}

\subsection{Find readout}

\subsubsection{Entropy optimisation}
\begin{outline}

    \1 Steering parameters: 
    
    \2 \verb|number_of_candidates| = 3 // Maximum number of candidates this sub-stage can create.
    \item \verb|seeding| = 15 // As above.
    \item \verb|iterations| = 30 // Number of total measurements taken by the Bayesian optimisation.

     \item \verb|freq_vs.minimum| = 2.6e9 // Minimum driving frequency $f_\text{MW}$ used in Bayesian optimisation, in Hz.
    \item \verb|freq_vs.maximum| = 2.9e9 // Maximum driving frequency $f_\text{MW}$ used in Bayesian optimisation, in Hz.
    \item \verb|burst_time_ns.minimum| = 2 // Minimum burst time $t_\text{burst}$ used in Bayesian optimisation, in ns.
    \item \verb|burst_time_ns.maximum| = 16 // Minimum burst time $t_\text{burst}$ used in Bayesian optimisation, in ns.

    \1 Measurement parameters:
    
    \2 \verb|magnetic_field| = 0.1  // As above.
    \item \verb|resolution| = 0.00075 // As above.
    \item \verb|padding| = 0.005 // As above.
    \item \verb|wait_time_slow_axis| = 0.5 // As above.
    \item \verb|wait_time_fast_axis| = 0.051 // As above.
    \item \verb|lockin_tc| = 1 // Time constant of lock-in amplifier, in s.
      \item \verb|field_setpoint.start| = 0.0 // Minimum magnetic field, in T. 
    \item \verb|field_setpoint.stop| = 0.4 // Maximum magnetic field, in T. 

    \item \verb|field_setpoint.num_points| = 300 // Number of points in magnetic field axis. 
    \item \verb|field_setpoint.delay| = 1 // Delay parameter, as above.

    \1 Analysis parameters:
    
    \2 \verb|segmentation_upscaling_res| = 2 // As above.
    \item \verb|relative_min_area| = 0.001 // As above.
    \item \verb|thr_method| = 'triangle' // As above.

\end{outline}

\subsubsection{Resonance confirmation}
\begin{outline}
    \1 Measurement parameters:
    
   \2 \verb|magnetic_field_window| = 0.1 // Window symmetric around the assumed peak, in T.
    \item \verb|resolution_magnet| = 0.001 // Resolution of scan, in T.

    \item \verb|wait_time| = 2.5 // Delay for measurement, needs to be longer than the lock-in time constant.

    \item \verb|lockin_tc| = 2 // As above.

    \1 Analysis parameters:
    
    \2 \verb|prominence| = 0.9 // Minimum prominence of peaks.
    \item \verb|sigma| = 1 // Gaussian smoothing factor.
    \item \verb|peak_offset_tolerance| = 0.025 // Peaks with a maximum offset of this parameter from the assumed position are accepted.

\end{outline}

\subsubsection{Spectroscopy}
\begin{outline}
    \1 Measurement parameters:
    
    \2 \verb|min_magnetic_field| = 0 // Start of magnetic field, in T.
    \item \verb|max_magnetic_field| = 0.5 // End of magnetic field, in T.
    \item \verb|resolution_magnet| = 0.005 // Resolution of magnetic field, in T.

    \item \verb|min_freq_vs| = 2e9 // Start of driving frequency $f_\text{MW}$, in Hz.
    \item \verb|max_freq_vs| = 3e9 // End of driving frequency $f_\text{MW}$, in Hz.

    \item \verb|resolution_freq| = 0.5e8 // Resolution of driving frequency $f_\text{MW}$, in Hz.

    \item \verb|extra_wait_time_slow_axis| = 1 // As above.
    \item \verb|wait_time_fast_axi|s = 1.1 // As above. 

    \item \verb|lockin_tc| = 1 // As above.

\end{outline}

\subsubsection{Rabi chevron}
\begin{outline}

        \1 Measurement parameters:
    \2 \verb|magnetic_field_window_multiplier| = 2 // The window is determined by the width of the peak that resonance confirmation (Stage 4b) takes and multiplied with this factor.
    \item \verb|n_px_magnet| = 40 // Number of points in the magnetic field axis. 

    \item \verb|resolution_burst_time| = 1e-9 // Resolution of the burst time $t_\text{burst}$ axis, in s.

    \item \verb|min_burst_time| = 0  // Minimum of the burst time $t_\text{burst}$, in s.
    \item \verb|max_burst_time| = 45e-9 // Maximum of the burst time $t_\text{burst}$, in s.
    \item \verb|dead_burst_time| = 10e-9 // The total length of the pulse is twice the maximum $t_\text{burst}$, plus this factor, in s.

    \item \verb|extra_wait_time_slow_axis| = 6 // As above. Needs to be significantly larger than the lock-in time constant to avoid spill-over effects.
    \item \verb|wait_time_fast_axis| = 2.5 // As above.

    \item \verb|lockin_tc| = 2 // As above.
\end{outline}  

\subsubsection{Rabi oscillations}
\begin{outline}
    \1 Steering parameters:
    \2 \verb|n_repetitions| = 5 // Number of repetitions of the same Rabi oscillation measurement.

    \1 Measurement parameters:
    
    \2 \verb|resolution_burst_time| = 0.5e-9 // As above.

    \item \verb|min_burst_time| = 0 // As above.
    \item \verb|max_burst_time| = 60e-9 // As above.
    \item \verb|dead_burst_time| = 10e-9 // As above.

    \item \verb|lockin_tc| = 2 // As above.

\end{outline}

\clearpage

\section{Efficient measurement algorithm}

\begin{figure}[htbp]
\centering
  \includegraphics[width=0.4\linewidth]{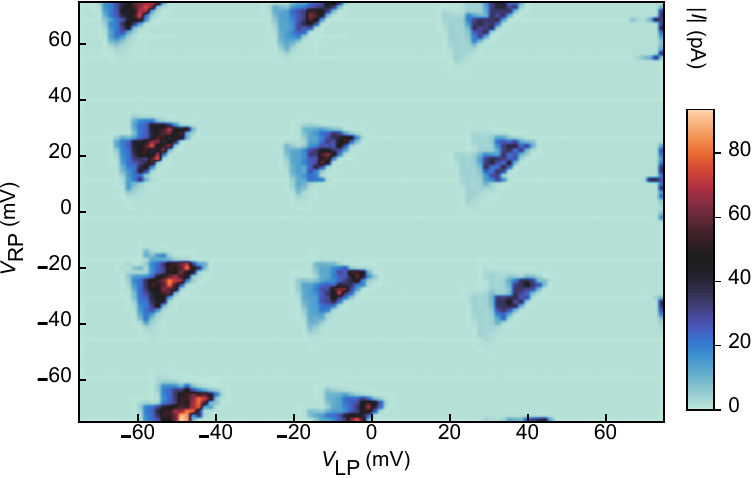}

  \caption{ \textbf{Examples of measurements taken with the efficient measurement algorithm.} Areas where no measurement have been taken have been filled with the threshold value.}
  \label{fig:snake_scan_example}
\end{figure}
Taking multiple charge transition stability diagram measurements is notably time-consuming, largely due to the predominance of featureless areas, as bias triangles are typically embedded within a skewed rectangular pattern. To address this, we have developed an efficient measurement algorithm. By rephrasing the measurement of bias triangles as an image processing task, whereby the goal is to determine a contour that traces the outline of a bias triangle, we were able to reduce the measurement time to 33$\%$ of a brute force scan. 

For a binary image represented by a matrix composed entirely of zeros and ones, the perimeter of any grouping of non-zero elements that form a contiguous region is known as a \textit{contour}. The Moore-Neighbour contour tracing algorithm provides a method of evaluating a complete contour given a starting point within the contour \cite{pavlidis2012algorithms}. The Moore-Neighbour contour achieves this by only ever examining pixels adjacent to a previously examined pixel. As the location of pixels corresponds to plunger gate voltages, measurements of well separated pixels are both costly in time and present a risk of introducing noise such as switches. Once the edge of a bias triangle has been identified, its contour can therefore be quickly measured with minimal overhead from the device. After the contour has been evaluated, each pixel inside the contour can be measured sequentially to complete the bias triangle.

To construct a binary image from a series of measured current values at differing gate voltages, a threshold must be determined. Current values above and below this threshold are considered to be ones and zeros in the binary image, respectively. The threshold can either be set manually, using prior knowledge of the system, or determined on-the-fly. 
To provide a fully automated system we took a calibration scan using a sparse sampling and evaluated the median absolute deviation threshold from these measured points
. The sampling routine was a so-called \textit{snake scan}, where measurements are performed horizontally left-to-right until a boundary of the measurement region is reached, then proceed vertically for a fixed length and continue horizontally in the opposite direction until the entire image has been covered.  

With a threshold determined, a second sparse sampling across the measurement region is performed.  The Moore-Neighbour contour tracing routine is triggered on any measurements that exceed the threshold, followed by a routine to measure the inside of the contour. This process systematically captures a complete pair of bias triangles. Post completion, the scan resumes until the next cluster necessitates flood filling. In order to minimize wasted measurements, current values were cached. The second sampling of the routine also used a snake scan to explore the measurement region, however it was offset vertically compared to the original to maximise the chance of encountering a bias triangle. 

Bias triangles are organised on a skewed grid pattern. The bias triangles evaluated in the second sparse sampling stage can be used to fit a skewed rectangular grid and infer the location of any missing bias triangles. Triangles can be missed by the initial sparse sampling if they reside between the horizontal lines of the two snake scans. A skewed grid can be represented by two vectors that describe the spatial separation between points on the grid, and the location of one grid point. These parameters were determined by minimising the total distance between the barycenter of all contours evaluated in the second sampling stage and points on the fitted grid. After fitting, each point on the grid within the measurement region that did not have a bias triangle was measured sequentially. 

Employing this method has proven to significantly streamline the process, cutting down the measurement time by approximately two-thirds. For the hyperparameters as reported above, the measurement of a 100 by 100 point stability diagram takes 14.5 min $\pm$ 3.0 min with this efficient measurement algorithm, and 43.7 min $\pm$ 0.1 min with a conventional grid scan.

%
%
\clearpage

\section{Search tree examples}
In Fig.~1b of the main text, we show an illustrative example of a search tree. Here, we show the search trees that were actually constructed for the longest and shortest runs in our experiments.

\begin{figure}[htbp]
\centering
  \includegraphics[width=1\linewidth]{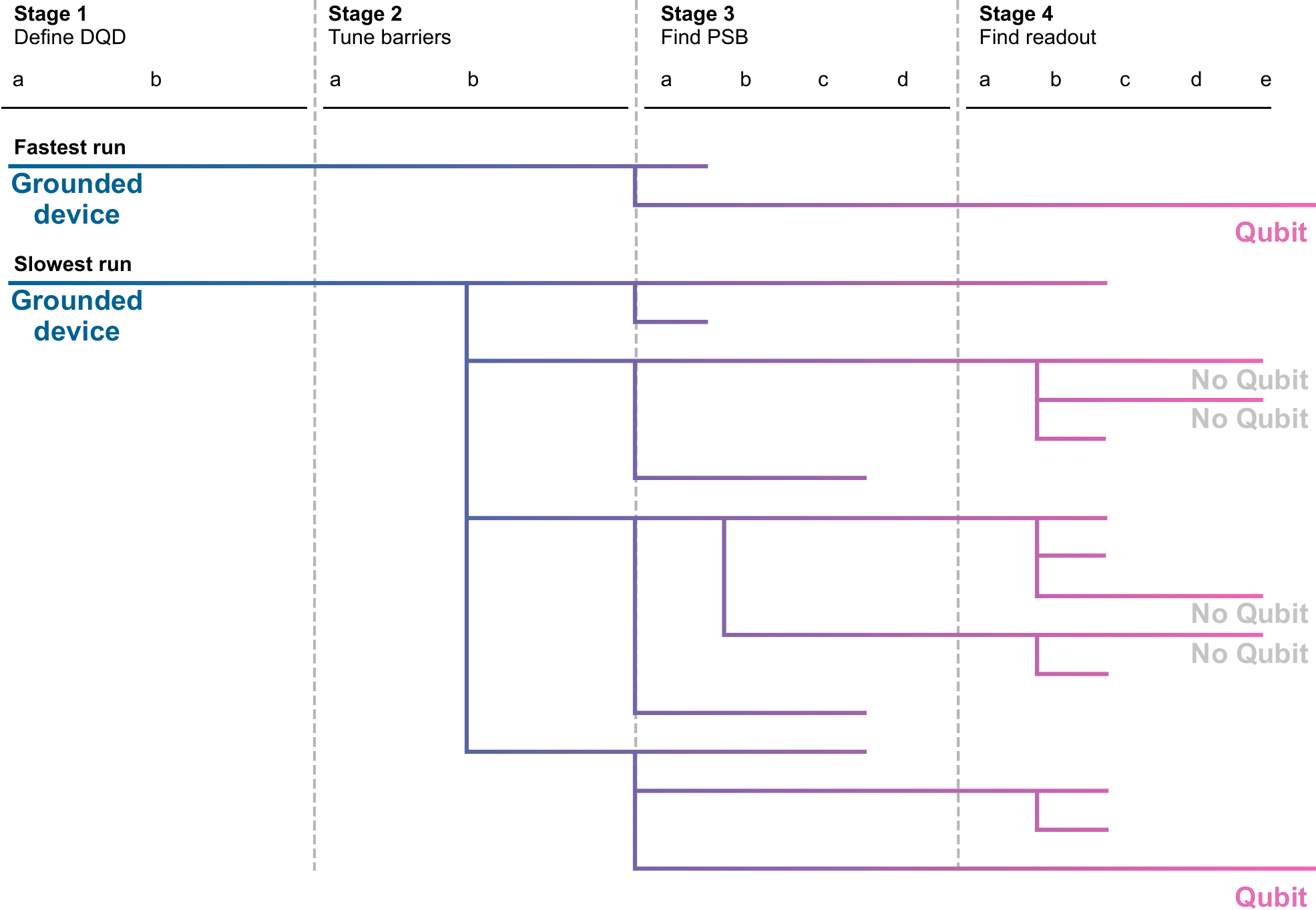}

  \caption{ \textbf{Examples of search trees from full runs.} The fastest run only has two branches and then successfully found a qubit. The slowest run explored much more, with several branches reaching all the way to qubit measurements. However, only the last branch shows conclusive qubit signatures. We rejected the first tries as noise.}
  \label{fig:tree_examples}
\end{figure}

\newpage

\clearpage

\section{Modular framework}
We implemented several design choices to standardize the framework across all stages, achieving a cohesive and modular architecture. Each stage exhibits these common characteristics:

\subsection{Stage structure}

\begin{enumerate}
    \item Integration with QCoDeS \cite{qcodes}: All stages have access to the station object of QCoDeS, allowing each stage to take measurements and change experimental parameters.

    \item Data management: A data access object manages (in addition to the QCoDeS database) costum data saving, such as the structure of the tree that was created so far, and automated documentation of the run.

    \item Hyperparameter configuration: Each stage possesses specifically tailored hyper-parameters to fulfill its requirements, as detailed in the Section \ref{supp:hyperparameters}.
    
    \item Candidate management: A list of candidates that were passed to a stage and that are sent off to another stage is kept.
    
\end{enumerate}

\subsubsection{Functions}
\begin{enumerate}

\item Investigation function: Stages are primarily invoked through an \verb|investigate| function, managing candidate lists, orchestrating measurements and data analysis, and forwarding candidates to the subsequent stage.

\item Experimental setup: A \verb|prepare_experiment| function sets up the experimental parameters as needed, for example, setting certain voltages, ramping the magnet to a starting position, or stopping the AWG from outputting a pulse sequence.

\item Experiment execution: The function \verb|perform_experiment| unction conducts the stage-specific measurements.

\item Data analysis: \verb|determine_candidate| function analyses the acquired data to assemble a viable set of candidates for further exploration.

\end{enumerate}

\subsection{Candidates}

\begin{enumerate}

    \item Data association: Once a stage has taken data relating to a specific candidate, it will keep a note of the global unique identifier (GUID) that is recorded in the QCoDeS database.
    
    \item Parameter storage: Critical parameter information is stored flexibly in a dictionary format to adapt to various experimental scenarios.
    
    \item Metadata storage: Candidates carry metadata, such as their position within the search tree.
    
    \item Stage timing: The duration required for each stage's process is recorded.
    
    \item Resulting candidates list: A distinct list is maintained for candidates resulting from the stage's analysis.
\end{enumerate}

\bibliography{reference}